\font\cmss=cmss10 
\font\cmsss=cmss10 at 7pt

\font\bigit=cmti10 scaled \magstep1

\def\unlockat{\catcode`\@=11}

\unlockat

\def\mdp{\bigskip\ndt}

% Sets global section and proposition numbers to zero
\global\newcount\secno \global\secno=0
\global\newcount\prono \global\prono=0
%%%%%%%%% Big sections
\def\newsec#1{\global\advance\secno by1\message{(\the\secno. #1)}
\global\subsecno=0\global\subsubsecno=0
\global\deno=0\global\teno=0
\eqnres@t\noindent
{\titlefont\the\secno. #1}
\writetoca{{\bf\secsym} { #1}}\par\nobreak\medskip\nobreak}
\global\newcount\subsecno \global\subsecno=0
%%%%%%%%%%%%%%%%Subsections %%%%%%%%%%%%%%%%%%%%%%%%%%%%%%%%%%%%%%%
\def\subsec#1{\global\advance\subsecno
by1\message{(\secsym\the\subsecno. #1)}
\ifnum\lastpenalty>9000\else\bigbreak\fi
\global\subsubsecno=0
\global\deno=0
\global\teno=0
%\eqnres@t
\noindent{\bf\secsym\the\subsecno. #1}
\writetoca{\bf \string\quad {\secsym\the\subsecno.} {\it  #1}}
\par\nobreak\medskip\nobreak}
\global\newcount\subsubsecno \global\subsubsecno=0
%%%%%%%%%%%%%%%%%%%%Subsubsections %%%%%%%%%%%%%%%%%%%%%%%%%%%%%
\def\subsubsec#1{\global\advance\subsubsecno by1
\message{(\secsym\the\subsecno.\the\subsubsecno. #1)}
\ifnum\lastpenalty>9000\else\bigbreak\fi
\noindent\quad{\bf \secsym\the\subsecno.\the\subsubsecno.}{\ \sl \ #1}
\writetoca{\string\qquad\bf { \secsym\the\subsecno.\the\subsubsecno.}{\sl  \ #1}}
\par\nobreak\medskip\nobreak}
%%% Definition

\global\newcount\deno \global\deno=0
\def\de#1{\global\advance\deno by1
\message{(\bf Definition\quad\secsym\the\subsecno.\the\deno #1)}
\ifnum\lastpenalty>9000\else\bigbreak\fi
\noindent{\bf Definition\quad\secsym\the\subsecno.\the\deno}{#1}
\writetoca{\string\qquad{\secsym\the\subsecno.\the\deno}{#1}}}
%%% Proposition

\global\newcount\prono \global\prono=0
\def\pro#1{\global\advance\prono by1
\message{(\bf Proposition\quad\secsym\the\subsecno.\the\prono %#1
)}
\ifnum\lastpenalty>9000\else\bigbreak\fi
\noindent{\bf Proposition\quad%\secsym\the\subsecno.
\the\prono\quad}{\ninepoint #1}
%\writetoca{\string\qquad{\secsym\the\subsecno.\the\prono}{#1}
}
%%% Theorem

\global\newcount\teno \global\prono=0
\def\te#1{\global\advance\teno by1
\message{(\bf Theorem\quad\secsym\the\subsecno.\the\teno #1)}
\ifnum\lastpenalty>9000\else\bigbreak\fi
\noindent{\bf Theorem\quad\secsym\the\subsecno.\the\teno}{#1}
\writetoca{\string\qquad{\secsym\the\subsecno.\the\teno}{#1}}}
%%%%%%%%%%%%
\def\subsubseclab#1{\DefWarn#1\xdef #1{\noexpand\hyperref{}{subsubsection}%
{\secsym\the\subsecno.\the\subsubsecno}%
{\secsym\the\subsecno.\the\subsubsecno}}%
\writedef{#1\leftbracket#1}\wrlabeL{#1=#1}}

\def\unredoffs{} \def\redoffs{\voffset=-.40truein\hoffset=-.40truein}
\def\speclscape{}

\newbox\leftpage \newdimen\fullhsize \newdimen\hstitle \newdimen\hsbody
\tolerance=1000\hfuzz=2pt

\catcode`\@=11
\def\bigans{b }
\def\answ{b }

\ifx\answ\bigans\message{(This will come out unreduced.}
\magnification=1200\unredoffs\baselineskip=16pt plus 2pt minus 1pt
\hsbody=\hsize \hstitle=\hsize

\else\message{(This will be reduced.} \let\l@r=L
\magnification=1200\baselineskip=16pt plus 2pt minus 1pt \vsize=7truein
\redoffs \hstitle=8truein\hsbody=4.75truein\fullhsize=10truein\hsize=\hsbody
\output={\ifnum\pageno=0

   \shipout\vbox{{\hsize\fullhsize\makeheadline}
     \hbox to \fullhsize{\hfill\pagebody\hfill}}\advancepageno
   \else
   \almostshipout{\leftline{\vbox{\pagebody\makefootline}}}\advancepageno
   \fi}
\def\almostshipout#1{\if L\l@r \count1=1 \message{[\the\count0.\the\count1]}
       \global\setbox\leftpage=#1 \global\let\l@r=R
  \else \count1=2
   \shipout\vbox{\speclscape{\hsize\fullhsize\makeheadline}
       \hbox to\fullhsize{\box\leftpage\hfil#1}}  \global\let\l@r=L\fi}
\fi

\def\Title#1#2{%\nopagenumbers
\abstractfont\hsize=\hstitle\rightline{#1}%
\vskip 5pt\centerline{\titlefont #2}\abstractfont\vskip .5in\pageno=0}
%

%&%

\def\draftmode{\message{ DRAFTMODE }\def\draftdate{{\rm preliminary draft:
\number\month/\number\day/\number\yearltd\ \ \hourmin}}%

\writelabels\baselineskip=20pt plus 2pt minus 2pt
  {\count255=\time\divide\count255 by 60 \xdef\hourmin{\number\count255}
   \multiply\count255 by-60\advance\count255 by\time
   \xdef\hourmin{\hourmin:\ifnum\count255<10 0\fi\the\count255}}}

\def\nolabels{\def\wrlabeL##1{}\def\eqlabeL##1{}\def\reflabeL##1{}}
\def\writelabels{\def\wrlabeL##1{\leavevmode\vadjust{\rlap{\smash%
{\line{{\escapechar=` \hfill\rlap{\sevenrm\hskip.03in\string##1}}}}}}}%
\def\eqlabeL##1{{\escapechar-1\rlap{\sevenrm\hskip.05in\string##1}}}%
\def\reflabeL##1{\noexpand\llap{\noexpand\sevenrm\string\string\string##1}}}
\nolabels
%

%\headline{\hfil\sl\dte}

\global\newcount\secno \global\secno=0
\global\newcount\meqno
\global\meqno=1
\def\eqnres@t{\xdef\secsym{\the\secno.}\global\meqno=1
\bigbreak\bigskip}
\def\sequentialequations{\def\eqnres@t{\bigbreak}}
\def\appendix#1#2{\vfill\eject\global\meqno=1\global\subsecno=0\xdef\secsym{\hbox{#1.}}
\bigbreak\bigskip\noindent{\bf Appendix #1. #2}\message{(#1. #2)}
\writetoca{Appendix {#1.} {#2}}\par\nobreak\medskip\nobreak}

\def\eqnn#1{\xdef #1{(\secsym\the\meqno)}\writedef{#1\leftbracket#1}%
\global\advance\meqno by1\wrlabeL#1}
\def\eqna#1{\xdef #1##1{\hbox{$(\secsym\the\meqno##1)$}}
\writedef{#1\numbersign1\leftbracket#1{\numbersign1}}%
\global\advance\meqno by1\wrlabeL{#1$\{\}$}}
\def\eqn#1#2{\xdef #1{(\secsym\the\meqno)}\writedef{#1\leftbracket#1}%
\global\advance\meqno by1$$#2\eqno#1\eqlabeL#1$$}

\newskip\footskip\footskip14pt plus 1pt minus 1pt

\def\footnotefont{\ninepoint}\def\f@t#1{\footnotefont #1\@foot}
\def\f@@t{\baselineskip\footskip\bgroup\footnotefont\aftergroup\@foot\let\next}
\setbox\strutbox=\hbox{\vrule height9.5pt depth4.5pt width0pt}
\global\newcount\ftno \global\ftno=0
\def\foot{\global\advance\ftno by1\footnote{$^{\the\ftno}$}}

\newwrite\ftfile
\def\footend{\def\foot{\global\advance\ftno by1\chardef\wfile=\ftfile
$^{\the\ftno}$\ifnum\ftno=1\immediate\openout\ftfile=foots.tmp\fi%
\immediate\write\ftfile{\noexpand\smallskip%
\noexpand\item{f\the\ftno:\ }\pctsign}\findarg}%
\def\footatend{\vfill\eject\immediate\closeout\ftfile{\parindent=20pt
\centerline{\bf Footnotes}\nobreak\bigskip\input foots.tmp }}}
\def\footatend{}

\global\newcount\refno \global\refno=1
\newwrite\rfile
\def\ref{[\the\refno]\nref}
\def\nref#1{\xdef#1{[\the\refno]}\writedef{#1\leftbracket#1}%
\ifnum\refno=1\immediate\openout\rfile=refs.tmp\fi \global\advance\refno
by1\chardef\wfile=\rfile\immediate \write\rfile{\noexpand\item{#1\
}\reflabeL{#1\hskip.31in}\pctsign}\findarg}

\def\findarg#1#{\begingroup\obeylines\newlinechar=`\^^M\pass@rg}
{\obeylines\gdef\pass@rg#1{\writ@line\relax #1^^M\hbox{}^^M}%
\gdef\writ@line#1^^M{\expandafter\toks0\expandafter{\striprel@x #1}%
\edef\next{\the\toks0}\ifx\next\em@rk\let\next=\endgroup\else\ifx\next\empty%
\else\immediate\write\wfile{\the\toks0}\fi\let\next=\writ@line\fi\next\relax}}
\def\striprel@x#1{} \def\em@rk{\hbox{}}
\def\lref{\begingroup\obeylines\lr@f}
\def\lr@f#1#2{\gdef#1{\ref#1{#2}}\endgroup\unskip}
\def\semi{;\hfil\break}
\def\addref#1{\immediate\write\rfile{\noexpand\item{}#1}}

\def\startrefs#1{\immediate\openout\rfile=refs.tmp\refno=#1}
\def\xref{\expandafter\xr@f}\def\xr@f[#1]{#1}
\def\refs#1{\count255=1[\r@fs #1{\hbox{}}]}
\def\r@fs#1{\ifx\und@fined#1\message{reflabel \string#1 is undefined.}%
\nref#1{need to supply reference \string#1.}\fi%
\vphantom{\hphantom{#1}}\edef\next{#1}\ifx\next\em@rk\def\next{}%
\else\ifx\next#1\ifodd\count255\relax\xref#1\count255=0\fi%
\else#1\count255=1\fi\let\next=\r@fs\fi\next}

%%%%%%%TABLE OF CONTENT%%%%%%%%%%%%%

\def\writetoc{\immediate\openout\tfile=BAtoc.tmp
    \def\writetoca##1{{\edef\next{\write\tfile{\noindent  ##1
    \string\leaderfill {\noexpand\number\pageno} \par}}\next}}}

%       and this lists table of contents on second pass

%
\catcode`\@=12 % at signs are no longer letters
%
%   Unpleasantness in calling in abstract and title fonts
\edef\tfontsize{\ifx\answ\bigans scaled\magstep3\else scaled\magstep4\fi}
\font\titlerm=cmr10 \tfontsize \font\titlerms=cmr7 \tfontsize
\font\titlermss=cmr5 \tfontsize \font\titlei=cmmi10 \tfontsize
\font\titleis=cmmi7 \tfontsize \font\titleiss=cmmi5 \tfontsize
\font\titlesy=cmsy10 \tfontsize \font\titlesys=cmsy7 \tfontsize
\font\titlesyss=cmsy5 \tfontsize \font\titleit=cmti10 \tfontsize
\skewchar\titlei='177 \skewchar\titleis='177 \skewchar\titleiss='177
\skewchar\titlesy='60 \skewchar\titlesys='60 \skewchar\titlesyss='60
\def\titlefont{\def\rm{\fam0\titlerm}% switch to title font
\textfont0=\titlerm \scriptfont0=\titlerms \scriptscriptfont0=\titlermss
\textfont1=\titlei \scriptfont1=\titleis \scriptscriptfont1=\titleiss
\textfont2=\titlesy \scriptfont2=\titlesys \scriptscriptfont2=\titlesyss
\textfont\itfam=\titleit
\def\it{\fam\itfam\titleit}\rm}
\font\authorfont=cmcsc10 \ifx\answ\bigans\else scaled\magstep1\fi
\ifx\answ\bigans\def\abstractfont{\tenpoint}\else \font\abssl=cmsl10 scaled
\magstep1 \font\absrm=cmr10 scaled\magstep1 \font\absrms=cmr7
scaled\magstep1 \font\absrmss=cmr5 scaled\magstep1 \font\absi=cmmi10
scaled\magstep1 \font\absis=cmmi7 scaled\magstep1 \font\absiss=cmmi5
scaled\magstep1 \font\abssy=cmsy10 scaled\magstep1 \font\abssys=cmsy7
scaled\magstep1 \font\abssyss=cmsy5 scaled\magstep1 \font\absbf=cmbx10
scaled\magstep1 \skewchar\absi='177 \skewchar\absis='177
\skewchar\absiss='177 \skewchar\abssy='60 \skewchar\abssys='60
\skewchar\abssyss='60
\def\abstractfont{\def\rm{\fam0\absrm}% switch to abstract font
\textfont0=\absrm \scriptfont0=\absrms \scriptscriptfont0=\absrmss
\textfont1=\absi \scriptfont1=\absis \scriptscriptfont1=\absiss
\textfont2=\abssy \scriptfont2=\abssys \scriptscriptfont2=\abssyss
\textfont\itfam=\bigit \def\it{\fam\itfam\bigit}\def\footnotefont{\tenpoint}%
\textfont\slfam=\abssl \def\sl{\fam\slfam\abssl}%
\textfont\bffam=\absbf \def\bf{\fam\bffam\absbf}\rm}\fi
\def\tenpoint{\def\rm{\fam0\tenrm}% switch back to 10-point type
\textfont0=\tenrm \scriptfont0=\sevenrm \scriptscriptfont0=\fiverm
\textfont1=\teni  \scriptfont1=\seveni  \scriptscriptfont1=\fivei
\textfont2=\tensy \scriptfont2=\sevensy \scriptscriptfont2=\fivesy
\textfont\itfam=\tenit \def\it{\fam\itfam\tenit}\def\footnotefont{\ninepoint}%
\textfont\bffam=\tenbf
\def\bf{\fam\bffam\tenbf}\def\sl{\fam\slfam\tensl}\rm}
\font\ninerm=cmr9 \font\sixrm=cmr6 \font\ninei=cmmi9 \font\sixi=cmmi6
\font\ninesy=cmsy9 \font\sixsy=cmsy6 \font\ninebf=cmbx9 \font\nineit=cmti9
\font\ninesl=cmsl9 \skewchar\ninei='177 \skewchar\sixi='177
\skewchar\ninesy='60 \skewchar\sixsy='60
\def\ninepoint{\def\rm{\fam0\ninerm}% switch to footnote font
\textfont0=\ninerm \scriptfont0=\sixrm \scriptscriptfont0=\fiverm
\textfont1=\ninei \scriptfont1=\sixi \scriptscriptfont1=\fivei
\textfont2=\ninesy \scriptfont2=\sixsy \scriptscriptfont2=\fivesy
\textfont\itfam=\ninei \def\it{\fam\itfam\nineit}\def\sl{\fam\slfam\ninesl}%
\textfont\bffam=\ninebf \def\bf{\fam\bffam\ninebf}\rm}
%
%---------------------------------------------------------------------
%

\hyphenation{anom-aly anom-alies coun-ter-term coun-ter-terms}
\def\inv{^{\raise.15ex\hbox{${\scriptscriptstyle -}$}\kern-.05em 1}}

\def\Dsl{\,\raise.15ex\hbox{/}\mkern-13.5mu D} 
%this one can be subscripted
\def\dsl{\raise.15ex\hbox{/}\kern-.57em\partial}

\def\tr#1{\, {\rm tr}\, \left( #1 \right)}

 %pound sterling
\def\lspace{\ifx\answ\bigans{}\else\qquad\fi}
\def\lbspace{\ifx\answ\bigans{}\else\hskip-.2in\fi} % $$\lbspace...$$
\def\boxeqn#1{\vcenter{\vbox{\hrule\hbox{\vrule\kern3pt\vbox{\kern3pt
     \hbox{${\displaystyle #1}$}\kern3pt}\kern3pt\vrule}
    }}}
\def\mbox#1#2{\vcenter{\hrule \hbox{\vrule height#2in
         \kern#1in \vrule} \hrule}}  %e.g. \mbox{.1}{.1}

%%%%%%%%%%%%%%%%%%%%%%%%

\newwrite\ffile\global\newcount\figno \global\figno=1
\def\nfig#1{\xdef#1{fig.~\the\figno}%
\writedef{#1\leftbracket fig.\noexpand~\the\figno}%
\ifnum\figno=1\immediate\openout\ffile=figs.tmp\fi\chardef\wfile=\ffile%
\immediate\write\ffile{\noexpand\medskip\noexpand\item{Fig.\ \the\figno. }
\reflabeL{#1\hskip.55in}\pctsign}\global\advance\figno by1\findarg}
\def\xfig{\expandafter\xf@g}
\def\xf@g fig.\penalty\@M\ {}
\def\figs#1{figs.~\f@gs #1{\hbox{}}}
\def\f@gs#1{\edef\next{#1}\ifx\next\em@rk\def\next{}\else
\ifx\next#1\xfig #1\else#1\fi\let\next=\f@gs\fi\next}
\newwrite\lfile
{\escapechar-1\xdef\pctsign{\string\%}\xdef\leftbracket{\string\{}
\xdef\rightbracket{\string\}}\xdef\numbersign{\string\#}}

\def\writestop{\def\writestoppt{\immediate\write\lfile{\string\pageno%
\the\pageno\string\startrefs\leftbracket\the\refno\rightbracket%
\string\def\string\secsym\leftbracket\secsym\rightbracket%
\string\secno\the\secno\string\meqno\the\meqno}\immediate\closeout\lfile}}
\def\writestoppt{}\def\writedef#1{}
\def\seclab#1{\xdef #1{\the\secno}\writedef{#1\leftbracket#1}\wrlabeL{#1=#1}}
\def\subseclab#1{\xdef #1{\secsym\the\subsecno}%
\writedef{#1\leftbracket#1}\wrlabeL{#1=#1}}
\newwrite\tfile \def\writetoca#1{}
\def\leaderfill{\leaders\hbox to 1em{\hss.\hss}\hfill}

%%%%%% Tildes and hats%%%%%%%%%%%%%%%

\def\tilde{\widetilde}
\def\bar{\overline}
\def\hat{\widehat}

%%%%%%%%%%%%% Cech %%%%%%%%%%%%
\def\cech{${\rm C}^{\kern-6pt \vbox{\hbox{$\scriptscriptstyle\vee$}\kern2.5pt}}${\rm ech}}
\def\Cech{${\sl C}^{\kern-6pt \vbox{\hbox{$\scriptscriptstyle\vee$}\kern2.5pt}}${\sl ech}}

%%%%%%%%%%%% Greek %%%%%%%%%%%%

\def\a{{\alpha}}

\def\d{{\delta}}

\def\e{{\epsilon}}

\def\vf{{\varphi}}

\def\l{{\lambda}}
\def\m{{\mu}}
\def\n{{\nu}}

\def\s{{\sigma}}
\def\t{{\theta}}
\def\vt{{\vartheta}}

%%%%% Non-terms %%%%%

%%%%%%%%%%%% Derivatives  %%%%%%%%%%%
\def\p{\partial}
\def\pb{\bar{\partial}}

\def\inv{^{\raise.15ex\hbox{${\scriptscriptstyle -}$}\kern-.05em 1}}

\def\Dsl{\,\raise.15ex\hbox{/}\mkern-13.5mu D}
%this one can be subscripted
\def\dsl{\raise.15ex\hbox{/}\kern-.57em\partial}

%%%%%%%%%%% bold %%%%%%%%%%%%%%

\def\bA{{\bf A}}

\def\bb{{\bf b}}
\def\bC{{\bf C}}

\def\bg{{\bf g}}

\def\bi{{\bf i}}

\def\bj{{\bf j}}

\def\bK{{\bf K}}
\def\bk{{\bf k}}

\def\bl{{\bf l}}
\def\bM{{\bf M}}
\def\bm{{\bf m}}
\def\bN{{\bf N}}
\def\bn{{\bf n}}

\def\bP{{\bf P}}
\def\bQ{{\bf Q}}

\def\bR{{\bf R}}
\def\bs{{\bf s}}
\def\bS{{\bf S}}
\def\bT{{\bf T}}
\def\bt{{\bf t}}

\def\bV{{\bf V}}

\def\bX{{\bf X}}

\def\bZ{{\bf Z}}

%%%%%%%%%%% letters with bar %%%%%%%%

%%%%%%%%% underlined letters %%%%%%%%%%%

%%%%%%%%%%%%%%%  Rublenye bukvy   %%%%%%%%%%%%%%%%%
\def\IB{\relax\hbox{$\inbar\kern-.3em{\rm B}$}}

\def\ID{\relax\hbox{$\inbar\kern-.3em{\rm D}$}}
\def\IE{\relax\hbox{$\inbar\kern-.3em{\rm E}$}}
\def\IF{\relax\hbox{$\inbar\kern-.3em{\rm F}$}}
\def\IG{\relax\hbox{$\inbar\kern-.3em{\rm G}$}}
\def\IGa{\relax\hbox{${\rm I}\kern-.18em\Gamma$}}
\def\IH{\relax{\rm I\kern-.18em H}}
\def\IK{\relax{\rm I\kern-.18em K}}
\def\IL{\relax{\rm I\kern-.18em L}}
\def\IP{\relax{\rm I\kern-.18em P}}
\def\IQ{\relax{{\vrule height1.5ex width.8pt depth0pt}\kern-.28em  Q\kern -.24em{\vrule height0.7em width.8pt depth0pt}}}
\def\IR{\relax{\rm I\kern-.18em R}}
\def\IZ{\relax\ifmmode\mathchoice{
\hbox{\cmss Z\kern-.4em Z}}{\hbox{\cmss Z\kern-.4em Z}}
{\lower.9pt\hbox{\cmsss Z\kern-.4em Z}} {\lower1.2pt\hbox{\cmsss
Z\kern-.4em Z}} \else{\cmss Z\kern-.4em Z}\fi}
\def\II{\relax{\rm I\kern-.18em I}}

\def\ndt{{\noindent}}

%%%%%%%% Calligraphic letters  %%%%%%%%%%%%%

\def\CC{{\cal C}}
\def\CD{{\cal D}}

\def\CF{{\cal F}}

\def\CH{{\cal H}}

\def\CL{{\cal L}}
\def\CM{{\cal M}}
\def\CN{{\cal N}}
\def\CO{{\cal O}}

\def\CQ{{\cal Q}}
\def\CR{{\cal R}}

\def\CT{{\cal T}}

\def\CW{{\cal W}}

\def\CY{{\cal Y}}

%%%%%%%%%%% letters with bar %%%%%%%%

%%%%%%%%%% Math symbols %%%%%%%%%%%%%

\def\Det{{\rm Det}}

\def\HH{{\bf H}}

\def\lime{{\rm Lim}_{\kern -16pt \vbox{\kern6pt\hbox{$\scriptstyle{\e \to 0}$}}}}

\def\naiveq{\qquad =^{\kern-12pt \vbox{\hbox{$\scriptscriptstyle{\rm naive}$}\kern5pt}} \qquad}

%
%---------------------------------------------------------------------
%

\hyphenation{anom-aly anom-alies coun-ter-term coun-ter-terms}
\def\tr{\, {\rm tr}\, }

 %pound sterling
\def\lspace{\ifx\answ\bigans{}\else\qquad\fi}
\def\lbspace{\ifx\answ\bigans{}\else\hskip-.2in\fi} % $$\lbspace...$$
\def\boxeqn#1{\vcenter{\vbox{\hrule\hbox{\vrule\kern3pt\vbox{\kern3pt
      \hbox{${\displaystyle #1}$}\kern3pt}\kern3pt\vrule}\hrule}}}
\def\mbox#1#2{\vcenter{\hrule \hbox{\vrule height#2in
          \kern#1in \vrule} \hrule}}  %e.g. \mbox{.1}{.1}
%   matters of taste
%\def\tilde{\widetilde} \def\bar{\overline} \def\hat{\widehat}
%
% some sample definitions

\def\darr#1{\raise1.5ex\hbox{$\leftrightarrow$}\mkern-16.5mu #1}
 %pound sterling

%%%%%%% halfs %%%%%%%%%%%

\def\half{{\textstyle{1\over2}}} 
\def\ihalf{{\textstyle{i\over2}}}

%puts a small half in a displayed eqn
\def\roughly#1{\raise.3ex\hbox{$#1$\kern-.75em\lower1ex\hbox{$\sim$}}}

%%%%%%%% Lie algebras %%%%%%%%%%%

\def\inbar{\,\vrule height1.5ex width.4pt depth0pt}

%%%%%%%%%%% Macros for boxes %%%%%%%%%%%
\def\boxit#1{\vbox{\hrule\hbox{\vrule\kern8pt
\vbox{\hbox{\kern8pt}\hbox{\vbox{#1}}\hbox{\kern8pt}}
\kern8pt\vrule}\hrule}}
\def\mathboxit#1{\vbox{\hrule\hbox{\vrule\kern8pt\vbox{\kern8pt
\hbox{$\displaystyle #1$}\kern8pt}\kern8pt\vrule}\hrule}}

%%%%%%%%%%%%%%%%%%%%%%%%

%%%%% Latex %%%%%%%
\def\mathcal#1{{\cal #1}}

\def\frac#1#2{{{ #1 }\over{ #2 }}}
\def\frac1#1{{1\over{#1}}}

%%%%%%%%%%%%%%%%%%
%%%% Journals %%%%%%%%%
\def\np#1#2#3{Nucl. Phys.\ {\bf B #1}(#2)#3}
\def\cmp#1#2#3{Comm. Math. Phys.\ {\bf #1}(#2)#3}
\def\plb#1#2#3{Phys. Lett. \ {\bf B #1} (#2) #3}  
\def\jhep#1#2#3{JHEP \ {\bf #1}(#2)#3}

\def\prl#1#2#3{Phys. Rev. Lett. \ {\bf #1}(#2)#3}

%%%% Literature %%%%
\lref\hitchin{N.~Hitchin, ``Stable bundles and integrable systems'', Duke Math
{\bf 54}  (1987) 91-114}
\lref\hid{N.~Hitchin, ``The self-duality equations on a Riemann surface'',
Proc. London Math. Soc. {\bf 55} (1987) 59-126 }
\lref\atbotti{M.~Atiyah, R.~Bott, 
``The Yang-Mills Equations Over
Riemann Surfaces'', Philosophical Transactions of the Royal Society of London. Series {\bA}, Mathematical and Physical Sciences, Volume 308, Issue 1505, pp. 523-615}

\lref\gerasimovshatashvili{A.~Gerasimov, S.L.~Shatashvili, ``Higgs Bundles, Gauge Theories and Quantum Groups'', \cmp{277}{2008}{323-367}, arXiv:hep-th/0609024}

\lref\gstwo{A.~Gerasimov, S.L.~Shatashvili, ``Two-dimensional gauge theories and quantum integrable systems'', arXiv:0711.1472, in, {\it "From Hodge Theory to Integrability and TQFT: tt*-geometry"}, pp. 239-262,
R.~Donagi and K.~Wendland, Eds., Proc. of Symposia in
Pure Mathematics Vol. 78, AMS, 
Providence, Rhode Island}

\lref\gntd{A.~Gorsky, N.~Nekrasov, ``Hamiltonian systems of Calogero type and two dimensional Yang-Mills theory'', arXiv:hep-th/9304047, \np{414}{1994}{213-238}}

\lref\mnph{J.A.~Minahan, A.P.~Polychronakos, ``Interacting Fermion Systems from Two Dimensional QCD,'' 
\plb{326}{1994}{288-294}, arXiv:hep-th/9309044    \semi
~~~~``Equivalence of Two Dimensional QCD and the $c=1$ Matrix Model'', \plb{312}{1993}{155-165}, arXiv:hep-th/9303153   \semi
~~~~ ``Integrable Systems for Particles with Internal Degrees of Freedom,''
\plb{302}{1993}{265-270}, arXiv:hep-th/9206046}

    \lref\korepin{V.~Korepin, N.~Bogolyubov, A.~Izergin, ``Quantum Inverse Scattering Method and Correlation Functions'', Cambridge  Monographs on Mathematical Physics, Cambridge University Press, 1997}
\lref\dubrovintt{B.~Dubrovin, ``Geometry and Integrability of Topological-Antitopological Fusion'', arXiv:hep-th/9206037,   \cmp{152}{1993}{539-564} }

\lref\cutr{For the current situation see, ``Integrability in String and Gauge Theory'', Utrecht, August, 2008}
\lref\mz{J.~Minahan, K.~Zarembo, ``The Bethe-Ansatz for ${\CN}=4$ Super Yang-Mills, ''
arXiv:hep-th/0212208 , \jhep{0303}{2003}{013}}
\lref\vafacr{C.~Vafa, ``Topological Mirrors and Quantum Rings,'' in, {\it Essays on Mirror Manifolds}, ed. S.-T.~Yau (Intl.Press, 1992)}
\lref\vc{S.~Cecotti, C.~Vafa, ``Topological Anti-Topological Fusion'', \np{367}{1991}{359-461}}
\lref\gnell{A.~Gorsky, N.~Nekrasov, ``Elliptic Calogero-Moser system from two dimensional current algebra'', arXiv: hep-th/9401021}
\lref\gnthd{A.~Gorsky, N.~Nekrasov, ``Relativistic Calogero-Moser model as gauged WZW theory'', \np{436}{1995}{582-608}, arXiv:hep-th/9401017}

\lref\basshort{N.~Nekrasov and S.~L.~Shatashvili,
``Quantum integrability and supersymmetric vacua'', IHES-P/08/59, TCD-MATH-09-05, HMI-09-02, arXiv:hep-th/0901.4748}
\lref\bass{N.~Nekrasov, S.~Shatashvili, ``Supersymmetric vacua and quantum integrability,'' to appear}
\lref\nazarov{M.~Nazarov, V.~Tarasov,``On irreducibility of tensor products of Yangian modules associated with skew Young diagrams'', arXiv:math/0012039, Duke~Math.~J. {\bf 112} (2002), 343-378\semi
M.~Nazarov, V.~Tarasov, ``On Irreducibility of Tensor Products of Yangian Modules'',
arXiv:q-alg/9712004,
Internat.~Math.~Research~Notices (1998) 125-150\semi
M.~Nazarov, V.~Tarasov, ``Representations of Yangians with Gelfand-Zetlin Bases'',
arXiv:q-alg/9502008,  J.~Reine~Angew.~Math. {\bf 496} (1998) 181-212\semi
M.~Nazarov, V.~Tarasov, ``Yangians and Gelfand-Zetlin bases'', arXiv:hep-th/9302102, Publ.~Res.~Inst.~Math.~Sci. Kyoto {\bf 30} (1994) 459-478}  
  
\lref\WitDonagi{R.~ Donagi, E.~ Witten,
``Supersymmetric Yang-Mills Theory and
Integrable Systems'', hep-th/9510101, Nucl.Phys.{\bf B}460 (1996) 299-334}
\lref\Witfeb{E.~ Witten, ``Supersymmetric Yang-Mills Theory On A
Four-Manifold,'' J. Math. Phys. {\bf 35} (1994) 5101.}
 
\lref\WittenTwoD{E.~Witten, ``Two dimensional gauge theory revisited'', arXiv:hep-th/9204083}

\lref\gmmm{A.~Gorsky, A.~Marshakov, A.~Mironov, A.~Morozov, ``${\CN}=2$ Supersymmetric QCD and Integrable Spin Chains: Rational Case $N_f < 2N_c$'', arXiv: hep-th/9603140, \plb{380}{1996}{75-80}}
\lref\ggm{A.~Gorsky, S.~Gukov, A.~Mironov,``SUSY field theories, integrable systems and their stringy/brane origin -- II'', arXiv:hep-th/9710239,
\np{518}{1998}{689-713}\semi
A.~Gorsky, S.~Gukov, A.~Mironov,
``Multiscale ${\CN}=2$ SUSY field theories, integrable systems and their stringy/brane origin -- I
'', arXiv:hep-th/9707120, \np{517}{1998}{409-461} \semi
R.~Boels, J.~de Boer,
``Classical Spin Chains and Exact Three-dimensional Superpotentials'',
arXiv:hep-th/0411110}

\lref\BlThlgt{M.~ Blau and G.~ Thompson, ``Lectures on 2d Gauge
Theories: Topological Aspects and Path
Integral Techniques", hep-th/9310144.}

\lref\kolyaf{N.~Reshetikhin, ``The functional equation method in the theory of exactly soluble quantum systems'', ZhETF {\bf 84} (1983), 1190-1201 (in Russian) Sov. Phys. JETP {\bf 57} (1983), 691-696 (English Transl.)}

\lref\baxteriii{R.J.~Baxter, ``Partition Function of the Eight-Vertex Lattice Model'' , Ann. Phys. {\bf 70} (1972) 193-228\semi
``One Dimensional Anisotropic Heisenberg Chain'', Ann. Phys. {\bf 70} (1972) 323-337\semi ``Eight-Vertex Model in Lattice Statistics and One Dimensional Anisotropic Heisenberg Chain I, II, III'', Ann. Phys. {\bf 76} (1973) 1-24, 25-47, 48-71}

\lref\baekr{
A.~N.~Kirillov,
N.~Yu.~Reshetikhin, ``Representations of Yangians
and multiplicities of the inclusion of the irreducible components
of the tensor product of representations of simple Lie algebras'' ,
(Russian) Zap. Nauchn. Sem. Leningrad. Otdel. Mat. Inst. Steklov.
(LOMI) 160 (1987), Anal. Teor. Chisel i Teor. Funktsii. 8,
211--221, 301; translation in J.~Soviet~Math. {\bf 52} (1990), no. 3,
3156--3164}

\lref\wittgr{E.~ Witten, ``The Verlinde Algebra And The Cohomology Of
The Grassmannian'',  hep-th/9312104}

\lref\niksw{N.~Nekrasov, ``Seiberg-Witten prepotential
from instanton calculus,'' arXiv:hep-th/0206161,  arXiv:hep-th/0306211}

\lref\nikokounkov{N.~Nekrasov, A.~Okounkov, ``Seiberg-Witten theory and random partitions,''
 arXiv:hep-th/0306238}

\lref\baxter{R.~Baxter, ``Exactly solved models in statistical mechanics'', London, Academic Press, 1982}

\lref\issues{A.~Losev, N.~Nekrasov, S.~Shatashvili, ``Issues in topological gauge theory'', \np{534}{1998}{549-611}, 
arXiv:hep-th/9711108}

\lref\gerasimov{A.~Gerasimov, ``Localization in
GWZW and Verlinde formula,'' hepth/9305090}

\lref\btverlinde{M.~ Blau, G.~Thomson,
``Derivation of the Verlinde Formula from Chern-Simons Theory and the
$G/G$ model,'' \np{408}{1993}{345-390} }

\lref\gyangian{H.~Nakajima, ``Quiver varieties and finite dimensional representations of quantum affine algebras,'' 
arXiv:math/9912158}
\lref\mvyangian{M.~Varagnolo,  ``Quiver varieties and Yangians,'' arXiv:math/0005277}

\lref\yangyang{C.~N.~Yang, C.~P.~Yang, J.~Math.~Phys~{\bf 10} (1969) 1115}

\lref\nikfive{N.~Nekrasov, ``Five dimensional gauge theories and relativistic integrable systems'',
\np{531}{1998}{323-344},
arXiv:hep-th/9609219}

\lref\phases{E.~Witten, ``Phases of ${\CN}=2$ Theories in Two Dimensions",
Nucl. Phys. {\bf B403} (1993) 159, hep-th/9301042}

\lref\nekhol{N.~Nekrasov, ``Holomorphic bundles and many-body systems'', arXiv:hep-th/9503157, \cmp{180}{1996}{587-604} }  
    
\Title{ \vbox{\baselineskip12pt
\hbox{IHES-P/09/09}
\hbox{TCD-MATH-09-04}
\hbox{HMI-09-01}
\hbox{NSF-KITP-09-11}}}
{\vbox{
\bigskip
\bigskip
 \centerline{SUPERSYMMETRIC VACUA }
 \bigskip
 \centerline{AND}
 \bigskip
 \centerline{BETHE ANSATZ}
}}
\medskip
\centerline{\authorfont Nikita A.
Nekrasov\footnote{$^{a}$}{On leave of absence
from ITEP, Moscow, Russia}$^{,1}$,
and Samson L. Shatashvili$^{1,2,3}$}
\vskip 0.5cm
\centerline{\it $^{1}$ Institut des Hautes Etudes Scientifiques,
Bures-sur-Yvette, France}
\centerline{\it $^{2}$ Hamilton Mathematical Institute, Trinity College,
Dublin 2, Ireland}
\centerline{\it $^{3}$ School of Mathematics, Trinity College, Dublin 2, Ireland}
\vskip 0.1cm

%\draftmode
\bigskip
\ndt
{\ninepoint
This note is a short announcement of some results of a longer paper where the 
supersymmetric vacua of two dimensional ${\CN}=4$ gauge theories with matter, softly broken
by the twisted masses down to
${\CN}=2$, are shown to be in
one-to-one correspondence with the eigenstates of integrable spin chain Hamiltonians.  The Heisenberg
$SU(2)$ $XXX$ spin chain is mapped to the two dimensional $U(N)$ theory with fundamental
hypermultiplets, the $XXZ$ spin chain is mapped  to the analogous three dimensional super-Yang-Mills theory compactified on a circle,  the $XYZ$ spin chain and eight-vertex model  are related to the four dimensional theory compactified on ${\bT}^{2}$.
The correspondence extends to any spin group, representations, boundary conditions, and inhomogeneity, it includes Sinh-Gordon and non-linear Schr\"odinger models as well as the dynamical spin chains such as the Hubbard model. Compactifications of four dimensional 
${\CN}=2$ theories on a two-sphere lead to the
instanton-corrected Bethe equations. 
We propose a completely novel way for the Yangian, quantum affine, and elliptic algebras to act as a symmetry of a union of quantum field theories.}
\vfill\eject
\newsec{Introduction}

The dynamics of gauge theory is a subject of long history and the ever growing importance.

\mdp$\underline{\rm Gauge\ theories\ and \ many-body\ systems.}$ 
\bigskip
In the last fifteen years or so it has become clear that the gauge theory dynamics in the vacuum sector is related to that of the quantum many-body systems. A classic example is the equivalence of the
pure Yang-Mills theory with the $U(N)$ gauge group in two dimensons and the system of $N$ free non-relativistic fermions on a circle. The same theory embeds as a supersymmetric vacuum sector of a (deformation of) ${\CN}=2$ super-Yang-Mills theory in two dimensions.

The Ref. \ref\higgs{G.~Moore, N.~Nekrasov, S.~Shatashvili,
``Integration over the Higgs branches'', 
\cmp{209}{2000}{97-121}, arXiv:hep-th/9712241} found a less trivial example of the gauge theory/many-body correspondence. Namely, the results of \higgs\ imply that the vacua of a
certain supersymmetric two dimensional $U(N)$ gauge theory with massive adjoint matter are described by the solutions of Bethe equations for the quantum Nonlinear Schr\"odinger equation (NLS) in the $N$-particle sector. The model of \higgs\ describes the $U(1)$-equivariant intersection theory on the Hitchin's 
moduli space \hitchin,\hid, just as the pure Yang-Mills
theory describes
 the intersection theory on moduli space of flat connections on a two dimensional Riemann surface \atbotti. This subject was revived in \gerasimovshatashvili,\gstwo,  by showing that the natural interpretation of the results of \higgs\ is in terms of the equivalence of the vacua of the $U(N)$ Yang-Mills-Higgs theory in a sense of \gerasimovshatashvili\ and the energy eigenstates of the $N$-particle Yang system, i.e. a system of $N$ non-relativistic particles on a circle with delta-function interaction.  Furthermore, \gerasimovshatashvili,\gstwo\ suggested that such a correspondence should be a general property of a larger class of supersymmetric gauge theories in various spacetime 
 dimensions\foot{Prior to \higgs\  a different connection to spin systems with long-range interaction appeared in two dimensional pure Yang-Mills theory with massive matter 
 \gntd,\mnph. The three dimensional lift of that gauge theory describes the relativistic interacting particles \gnthd,
while the four dimensional theories lead to elliptic generalizations
\gnell.}.

\bigskip
\ndt$\underline{\rm A\ dictionary.}$ 
\bigskip
We thus aim to formulate precisely in full generality the correspondence between the two dimensional ${\CN}=2$ supersymmetric gauge theories and quantum integrable systems. 

{}The ${\CN}=2$ supersymmetric theories have rich algebraic structure surviving quantum corrections \vc. In particular, there is a distinguished class of operators $\left( {\CO}_{A} \right)$, which commute with some nilpotent supercharge $\CQ$ of the supersymmetry algebra. These operators have no singularities in their operator product expansion and, when considered up to the $\CQ$-commutators,  form a (super)commutative ring, called the (twisted) chiral ring \vc,\vafacr. The supersymmetric vacua of the theory form a representation of that ring. The space of supersymmetric vacua is thus naturally identified with the space of states of a quantum integrable system, whose Hamiltonians are the generators of the (twisted)  chiral ring.
Our duality states that the spectrum of the quantum Hamiltonians coincides with the spectrum of the (twisted) chiral ring.
The nontrivial result of this paper and that of \basshort,\bass is that arguably all quantum integrable lattice models from the integrable systems textbooks correspond in this fashion to the ${\CN}=2$
supersymmetric {\it gauge} theories, essentially also from the
(different) textbooks. More precisely, the gauge theories which correspond to the integrable spin chains and their limits (the non-linear 
Schr\"odinger equation and other systems encountered in \higgs,\gerasimovshatashvili,\gstwo\ being particular large spin limits thereof) are the softly broken
${\CN}=4$ theories. It is quite important that we are dealing here with the gauge theories, rather then the general $(2,2)$ models, since it is in the gauge theory context that the equations describing the supersymmetric vacua can be identified with Bethe equations of the integrable world.

It is known that the low energy
dynamics of the four dimensional ${\CN}=2$ supersymmetric
gauge theories is governed by the classical algebraic integrable systems \WitDonagi. Moreover, the natural
gauge theories lead to the integrable systems of Hitchin type \nikokounkov, which are equivalent to many-body systems \nekhol\ and conjecturally to spin chains
\gmmm,\ggm.

We emphasize, however, that the correspondence between
the  gauge theories and integrable models we discuss in the present paper, as well as in 
\higgs,\gerasimovshatashvili,\gstwo,\basshort,\bass\ is of a different nature.  The vacuum states we discuss presently are mapped to the quantum eigenstates of a different, quantum integrable system\foot{Another possible source of confusion is the emergence of the
Bethe ansatz and the spin chains in the ${\CN}=4$ supersymmetric
gauge theory in four dimensions.
In the work \mz\ and its further developments \cutr\  the anomalous dimensions of  local operators of the ${\CN}=4$ supersymmetric Yang-Mills theory
are shown (to a certain loop order in perturbation theory) to be the eigenvalues of
some spin chain Hamiltonian. The
gauge theory is studied in the 't Hooft large $N$ limit. In our story the gauge theory has less supersymmetry, $N$ is finite, and the operators we consider are from the chiral ring, i.e. their conformal dimensions are not corrected
quantum mechanically. Our goal is
to determine their vacuum expectation values.}.

The gauge theories we study in two dimensions, as well as their string theory realizations,  have a natural lift to three and four dimensions,
while keeping the same number of supersymmetries, modulo certain
anomalies. Indeed, the ${\CN}=2$
super-Yang-Mills theory in two dimensions is a dimensional reduction of the ${\CN}=1$ four dimensional Yang-Mills theory
(this fact is useful in
the superspace formulation
of the theory \phases).
Instead of the dimensional reduction one can study the compactification on a two dimensional torus ${\bT}^{2}$. The theory obtained in this way looks two dimensional macroscopically, yet its effective low energy dynamics gets
corrected by the loops of the Kaluza-Klein modes (the examples of these
corrections in the analogous
compactifications from five to four dimensions can be found in \nikfive). This is seen, for example, in the geometry of the (classical) moduli space of vacua, which
is compact for the theory obtained by compactification from four to two dimensions (being isomorphic to the moduli space ${\rm Bun}_{G}$ of the semi-stable holomorphic $G_{\bC}$-bundles  on elliptic curve), and is non-compact in the dimensionally reduced theory. Quantum mechanically, though, the geometry of the moduli space of vacua is more complicated, in particular it will acquire many components. The twisted superpotential is a meromorphic function on the moduli space.
We show that the critical points of this function determine the Bethe roots of the anisotropic spin chain, the $XYZ$ magnet. Its $XXZ$ limit will
be mapped to the
three dimensional gauge theory compactified on a circle. We thus get a satisfying picture of the elliptic, trigonometric, and rational theories  corresponding to the four dimensional, three dimensional  and the two dimensional theories respectively.

Our duality between the gauge theories and the quantum integrable systems can be used to enrich both subjects.

\bigskip
\ndt$\underline{\rm A\ longer\ version.}$ 
\bigskip
{}This note is a shortened version of \bass, where we give all the details covering the correspondence between vacuum structure of supersymmetric gauge theories and quantum integrable models from all perspectives, including the sring theory realization. Here we just mention that the
guiding equations for the supersymmetric vacua for the two, three, and four dimensional models
(compactified on the tori of appropriate dimension) can be summarised as:
\eqn\main{
{\exp} \left( {\p {\tilde W}^{\rm eff}}(\sigma) \over {\p \s}^{i}  \right) =  1}
where ${\tilde W}^{\rm eff}(\sigma)$ is the effective twisted superpotential, while $\sigma_i$ are the  eigenvalues of the complex scalar in the vector multiplet.  It is this equation that coincides with the Bethe equation determining the exact spectrum of a quantum integrable system. In this correspondence ${\tilde W}^{\rm eff}( {\s} )$ coincides with Yang-Yang function $Y( {\l} )$ 
(${\l}_i$ denoting the rapidities) generating the Bethe roots in quantum integrable systems \yangyang:
 \eqn\mainone{\mathboxit{\eqalign{ Y ({\l} )
\, & \, \leftrightarrow \, {\tilde W}^{\rm eff} ({\s}) \cr
  {\l} \, &\, \leftrightarrow {\s} \cr}}}
We identify these quantum integrable systems in all our examples and study the consequences.
 In \bass\ the Hamiltonians of the quantum integrable system are identified with the operators
of quantum multiplication in the equivariant cohomology of the
hyperk\"ahler quotients, corresponding to the Higgs branches of
our gauge theories. In particular, the length $L$ inhomogeneous $XXX_{1\over 2}$ chain (with  all local spins equal to $\half$) corresponds to the equivariant quantum cohomology of the cotangent bundle $T^{*} Gr(N, L)$ to the Grassmanian $Gr(N, L)$.
This result complements nicely the construction
of H.~Nakajima and others of the action of the Yangians \gyangian,\mvyangian\ and quantum affine algebras on the classical cohomology and K-theory respectively of certain quiver varieties. Next, \bass\ applies these results to the two dimensional topological field theories. We discuss various twists of our supersymmetric gauge theories. The correlation functions of the
chiral ring operators map to the equivariant intersection indices
on the moduli spaces of solutions to various versions of the
two dimensional vortex equations, with what is mathematically called the Higgs fields taking values in various line bundles (in the case of Hitchin equations the Higgs field is valued in the canonical line bundle).
The main body of \bass\ has essentially shown that all known Bethe ansatz-soluble integrable systems are covered by our correspondence. However, there
are more supersymmetric gauge theories which lead
to the equations \main\  which can be viewed as the deformations of Bethe equations. For example, a four dimensional ${\CN}=2^{*}$ theory compactified on ${\bS}^{2}$
with a partial twist leads to a  deformation
of the non-linear Schr\"odinger system with interesting modular
properties (we devote last section of current paper to this example). Another interesting model related to the $D1-D5$ brane systems relates   the quantum cohomology of instanton moduli spaces and the Hilbert scheme
of points \ref\okpan{A.~Okounkov, R.~Pandharipande, ``Quantum cohomology of the Hilbert scheme of points in the plane'', arXiv:math/0411210} to the Bethe ansatz for the
yet unknown spin chains with the affine Lie algebras replacing the $su(2)$ of the Heisenberg spin chain.

The long paper \bass\ is also reviewed in detail in
\basshort. In particular, the details of the correspondence between the equivariant quantum cohomology of T$^{*}$Gr$(N,L)$ and the
inhomogeneous Heisenberg magnet can be found there. 

\bigskip

\noindent {\bf Acknowledgments.}  We thank V.~Bazhanov, G.~Dvali, L.~Faddeev, S.~Frolov, A.~Gorsky, K.~Hori, A.~N.~Kirillov,
V.~Korepin, B. McCoy, M.~Nazarov, A.~Niemi,  A.~Okounkov, E.~Rabinovici, N.~Reshetikhin, S.~J.~Rey,  L.~Takhtajan,  A.~Vainshtein and P. Wiegmann, and especially A.~Gerasimov and F.~ Smirnov,  for discussions. The results of this note, as well as those in \bass,  were presented at various conferences and workshops\foot{The IHES seminars and the theoretical physics conference dedicated to the 50th anniversary of IHES  (Bures-sur-Yvette, June 2007, April 2008, June 2008);
 the IAS Workshop on ``Gauge Theory and Representation Theory'' and the IAS seminar (Princeton, November 2007, 2008); the
YITP/RIMS conference ``30 Years of Mathematical Methods
in High Energy Physics
'' in honour of 60th anniversary of Prof. T.~Eguchi (Kyoto, March 2008); the London Mathematical Society lectures at Imperial College  (London, April 2008);  L.~Landau's 100th anniversary
theoretical physics conference (Chernogolovka, June 2008); Cargese Summer Institute (Cargese, June 2008); the Sixth Simons Workshop ``Strings, Geometry and the LHC'' (Stony Brook, July 2008);
the ENS summer institute (Paris, August 2008); the French-Japanese Scientific Forum ''Perspectives in mathematical sciences'', (Tokyo,
October 2008)} and we thank the organizers for the opportunity to present our results. We thank various agencies and institutions\foot{The RTN contract 005104 "ForcesUniverse" (NN and SS),  the ANR grants
ANR-06-BLAN-3$\_$137168 and ANR-05-BLAN-0029-01 (NN), the RFBR grants
RFFI 06-02-17382 and NSh-8065.2006.2 (NN),
 the NSF grant No. PHY05-51164 (NN),  the SFI grants 05/RFP/MAT0036, 08/RFP/MTH1546 (SS) and the Hamilton Mathematics Institute TCD (SS).
Part of research was done while NN visited NHETC at Rutgers University in 2006, Physics and Mathematics Departments of Princeton University in 2007, Simons Center at the Stony Brook University in 2008, KITP at the UC Santa Barbara in 2009, while SSh visited IAS in Princeton in 2007, CERN in 2007 and 2008, Ludwig-Maximilians University in Munich in 2007 and IAS in Jerusalem in 2008.} for supporting this research.

\newsec{The gauge theory}

Here we give a brief review of the relevant gauge theories.

\subsec{Gauge theories with four supercharges}

We study two dimensional ${\CN} = (2,2)$ supersymmetric gauge theory
 with some matter. The matter fields are generally in
the chiral multiplets which we denote by the letters ${\bf Q}$, ${\tilde{\bf Q}}$, and ${\bf \Phi}$  (sometimes we use $\bX$ to denote matter fields without reference to their gauge representation type), the gauge fields are in the vector multiplet $\bf V$. We also use the twisted chiral multiplets $\bf\Sigma$, as e.g. the field strength ${\bf\Sigma} = {\CD}_{+}{\bar \CD}_{-}{\bV}$ is in the twisted chiral multiplet.
\eqn\multiplet{\eqalign{& {\bV}={\t}^{-}
{\bar\t}^{-}(A_{0}-A_{1})+{\t}^{+}{\bar\t}^{+}(A_{0}+A_{1})-\sqrt{2}{\s}{\t}^{-}{\bar\t}^{+}
-\sqrt{2}{\bar\s}{\t}^{+}{\bar\t}^{-}+ \cr
&
+2i{\t}^{-}{\t}^{+}({\bar\t}^{-}{\bar\l}_{-} +
{\bar\t}^{+}{\bar\l}_{+}) + 2 i {\bar\t}^{+}{\bar \t}^{-}({\t}^{+}{\l}_{+} +{\t}^{-}{\l}_{-})+2{\t}^{-}{\t}^{+}{\bar \t}^{-}{\bar\t}^{+}H \ , \cr}}
where we use a notation $H$ for the auxiliary field (in most textbooks it is denoted by $D$).
\eqn\chirl{{\bX} = X ( y ) + \sqrt{2} \left( {\t}^{+} {\psi}_{+}(y)  + {\t}^{-} {\psi}_{-}(y)  \right) + {\t}^{+}{\t}^{-} F(y)}
where $$
y^{\pm} = x^{\pm} - i{\t}^{\pm}{\bar\t}^{\pm}\ ,
$$
and the twisted chiral multiplet $\bf\Sigma$:
\eqn\sig{{\bf\Sigma} =  {\s}({\tilde y}) +i\sqrt{2}\left( {\t}^{+}{\bar \l}_{+}({\tilde y}) -{\bar \t}^{-}{\l}_{-}({\tilde y}) \right)  + \sqrt{2}{\t}^{+}{\bar \t}^{-} \left(  H ({\tilde y}) -i F_{01} \right)}
where $F_{01} = {\p}_{0}A_{1} - {\p}_{1}A_{0} + [A_{0}, A_{1}]$ is the gauge field strength, and
\eqn\ypmv{
{\tilde y}^{\pm} = x^{\pm} \mp i {\t}^{\pm}{\bar\t}^{\pm}}

\subsubsec{Lagrangians}

The action of the
corresponding two dimensional quantum field theory action has three types of terms - the $D$-terms, the $F$-terms and the twisted $F$-terms:
\eqn\dff{\eqalign{
D: \quad \quad & \quad \quad \int {\rm d}^{2}x\, {\rm d}^{4}{\theta}\ {\tr} \left( {\bf\Sigma}{\bf\bar \Sigma} \right) + {\bK} ( e^{{\bV}/2}\, {\bX}\ , \,{\bar {\bX}}\, e^{{\bV}/2})\cr
F:  \quad \quad &  \quad \quad \int {\rm d}^{2}x\, {\rm d}{\theta}^{+}{\rm d}{\t}^{-} \ W({\bX}) \, + \, {\rm c.c.}\cr
F^{\rm tw}: \quad \quad & \quad \quad \int {\rm d}^{2}x\, {\rm d}{\t}^{+}{\rm d}{\bar\t}^{-}\ {\tilde W}({\bf\Sigma}) \, + \, {\rm c.c.}\cr}}

\subsubsec{Global symmetries and twisted masses}

The typical ${\CN}=(2,2)$ gauge theory has the matter fields $\bX$ transforming in some
linear\foot{In \bass\ we also discuss the generalization where $\bX$ takes values in some non-linear space with the $G$-action} representation $\CR$ of the gauge group $G$.
Let us specify the decomposition of $\CR$ onto the irreducible representations of $G$:
\eqn\CRdc{{\CR} = \bigoplus_{\bi} \, {\bM}_{\bi} \otimes R_{\bi}}
where $R_{\bi}$ are the irreps of $G$, and ${\bM}_{\bi}$ are the multiplicity spaces.
The group
\eqn\glbgrp{H^{\rm max} = \times_{\bi} \, U({\bM}_{\bi})}
acts on $\CR$ and this action commutes with the gauge group action. The actual global symmetry
group $H$ of the theory may be smaller then \glbgrp : $H \subset H^{\rm max}$, as it has to preserve both $D$ and the $F$-terms in the action.

The theory we are interested in can be deformed by turning on the so-called twisted masses
${\tilde m}$ \ref\agf{L.~Alvarez-Gaume, D.Freedman, \cmp{91}{1983}{87}}, which belong to the complexification of the Lie algebra of the maximal torus of $H$:
\eqn\twmsss{{\tilde m} = \left( {\tilde m}_{\bi}  \right) \, , \, {\tilde m}_{\bi} \in {\rm End} \left( {\bM}_{\bi} \right) \cap H}
The superspace expression for the twisted mass term is \ref\sstwm{S.J.~Gates, \np{238}{1984}{349}},
\ref\sstwmii{S.J.~Gates, C.M.~Hull,  M.~Rocek, \np{248}{1984}{157}}:
\eqn\twmass{{\CL}_{\widetilde{\rm mass}} = \int {\rm d}^{4}{\theta}
\ {\tr}_{\CR} \, {\bX}^{\dagger} \left( \sum_{\bi} e^{{\tilde V}_{\bi}}\otimes {\rm Id}_{R_{\bi}}\right) {\bX} }
where
\eqn\tvm{{\tilde V}_{\bi} = {\tilde m}_{\bi}\, {\theta}_{+}{\bar\theta}_{-}}
The twisted masses which preserve the ${\CN}=4$ supersymmetry will be denoted by $\m$, and
the ones which break it down to
${\CN}=2$, by $u$.

When the twisted masses are turned on in  the generic fashion, the matter fields
are massive and can be integrated out. As a result, the theory becomes an effective pure ${\CN}=2$ gauge theory with an infinite
number of interaction terms in the Lagrangian, with the high derivative terms suppressed by the inverse masses of the fields we integrated out. Of all these terms the $F$-terms, i.e. the effective superpotential, or the twisted $F$-terms, i.e. the effective twisted superpotential, can be computed exactly. In fact, these terms only receive one-loop contributions. Let $\tilde\bm$ denote collectively the set of the twisted masses of the fields we are
integrating out.
We get:
\eqn\effsp{\widetilde{W^{\rm eff}}_{\rm matter}({\sigma}) = \sum_{\bb} 2{\pi}i\, t_{\bb} {\tr}_{\bb}{\s} + {\tr}_{\CR} \left( {\sigma} + {\tilde\bm} \right) \left( {\rm log}\left( {\sigma} + {\tilde\bm} \right)  - 1 \right)}
where for each $U(1)$ factor in $G$ we have introduced a Fayet-Illiopoulos term which together with the corresponding theta-angle combine into a complex coupling $t_{\bb}$,
\eqn\cmplfi{
t_{\bb} =  {{\vt}_{\bb}\over 2\pi} + i r_{\bb} \ .}
The generator of the corresponding $U(1)$ factor in $G$
is denoted in \effsp\ by ${\tr}_{\bb}{\s}$.
We put the subscript ``matter'' in \effsp\ in order to stress the fact that it only includes the loops of the matter fields.

There are other massive fields which can be integrated out on the Coulomb branch. For example, the ${\bg}/{\bt}$-components of the vector multiplets (where $\bg$ denotes Lie  algebra  corresponding to Lie groups $G$ and $\bt$ is its Cartan sub-algebra), the $W$-bosons
and their superpartners. Their contribution to the effective twisted superpotential is rather simple:
\eqn\gsup{
{\tilde W}^{\rm eff}_{\rm gauge} = -
\sum_{{\a} \in {\Delta}} \, \langle
{\a} , {\s} \rangle \, \left[ \,
{\rm log} \, \langle
{\a} , {\s} \rangle \, - 1\, \right]
= - 2{\pi} i \, \langle {\rho} , {\s} \rangle}
where
\eqn\hlfsm{{\rho} = {\half} \sum_{{\a} \in {\Delta}_{+}} {\a}}
is half the sum  of the positive roots of $\bg$.
It may appear that the expression \gsup\ is inconsistent
with the gauge invariance, however the effective interaction \gsup\ is gauge invariant.
The total effective twisted superpotential is, therefore:
\eqn\wsuptot{{\tilde W}^{\rm eff}({\s}) =
{\tilde W}^{\rm eff}_{\rm matter} ({\s})+ {\tilde W}^{\rm eff}_{\rm gauge}({\s})}

\subsubsec{Superpotential deformations and twisted masses}

The supersymmetric field theories also have the superpotential deformations, which correspond to the $F$-terms in \dff. The  superpotential $W$ has to be a holomorphic gauge invariant
function of the chiral fields, such as ${\Phi}, Q, {\tilde Q}$.
It may be not invariant under the maximal symmetry group $H^{\rm max}$,
thus breaking it to a subgroup $H$ or completely. For example, the so-called {\it complex mass} of the fundamental and anti-fundamental fields \HH\ comes from
the superpotential $
W_{\rm complex \, mass}  = \sum_{a,b}m_{a}^{b} {\tilde Q}_{b}Q^{a}
$,
which breaks the $U(n_{\bf f}) \times U(n_{\overline{\bf f}})$
group down to $U(1)^{{\rm min}(n_{\bf f}, n_{\overline{\bf f}})}$.

{}In all cases discussed in this paper, in spacetime dimensions two, three and four, one can consider more sophisticated superpotentials,  involving  the fundamental, anti-fundamental, and adjoint chiral fields:
\eqn\wqfq{
W_{\tilde Q {\Phi}Q}  = \sum_{a,b} {\tilde Q}^{a}m_{a}^{b}(\Phi)Q_{b} = \sum_{a,b; s}\, m_{a; s}^{b} {\tilde Q}^{a}
{\Phi}^{2s}Q_{b}}
The case of most interest for us, that of the two dimensional ultraviolet finite theories corresponds to $n_{\bf f}= n_{\overline{\bf f}} = L$. In this case we will see later that equations describing supersymmetric vacua are linked to known quantum integrable lattice models.

\subsec{Examples}

There are two classes of examples: $a.)$ the asymptotically free theories and $b.)$ the  asymptotically conformal theories. The $a.)$ examples include the gauge theories which look at low energy as the ${\CN}=2$ sigma models with various K\"ahler target spaces: the complex projective space ${\bC\bP}^{L-1}$, the
Grassmanian ${\rm Gr}(N, L )$, or, more generally, the (partial) flag variery $F ( n_{1} , n_{2} , \ldots , n_{\bf r} , n_{{\bf r}+1} \equiv L)$. The $b.)$ examples can also be identified at the low energy level with the sigma models. These sigma models typically have the hyperk\"ahler target spaces, such as the cotangent bundles to the K\"ahler manifolds from the $a.)$ list. The $b.)$ examples turn out to include, via \main,  essentially all known quantum integrable models of statistical physics.

By taking an appropriate scaling limit one can get the $a.)$ models from the $b.)$ models. For example, the Grassmanian model (which is so extensively studied in \wittgr) is a limit of the $T^{*}{\rm Gr}(N,L)$ model in the limit where the twisted mass $u$ corresponding to the
rotations of the cotangent direction is sent to infinity, with the complexified K\"ahler class adjusted in such a way, that the effective mass scale
$
{\Lambda}_{\rm Gr} = u e^{2\pi i t \over L}
$
remains finite. This corresponds to a non-Hermitian deformation of the Heisenberg magnet
which is dual, via \main, to the original $T^{*}{\rm Gr}(N,L)$ theory.

The reason why the ultraviolet finiteness is so special in the relation to the quantum integrability has to do with the $S$-matrix nature of the Bethe equations which we identify with the
vacuum equation \main.

In this note we consider the $G = U(N)$ gauge group only.
Here we present the effective twisted superpotential \wsuptot\ for the main example of the $b.)$ class. There are many more examples presented in \bass.

\subsubsec{Two dimensions}

One can start with
the so-called ${\CN}=2^{*}$ theory.
It has
 ${\CR} = {\bg} \otimes {\bC}$, i.e. the adjoint chiral multiplet $\bf\Phi$. In the absence of the twisted mass term this is the ${\CN}=4$ theory, the dimensional reduction of the pure ${\CN}=2$ super-Yang-Mills from four dimensions. This theory has a global $U(1)$ symmetry, which rotates the adjoint chiral multiplet, e.g. ${\bf\Phi} \mapsto e^{i \vf} {\bf\Phi}$. We can turn on the corresponding twisted mass ${\tilde m} = i u$ which breaks ${\CN}=4$ to ${\CN}=2$ (the factor of $i$ is introduced for the later convenience).  The effective twisted superpotential for $G= U(N)$ is:
\eqn\effspadj{{\tilde W}^{\rm eff} ({\s}) \, = \,
\sum_{i,j=1}^{N} \left( {\s}_{i} - {\s}_j + i u \right)
\left( {\rm log}\left( {\s}_{i} - {\s}_j + i u \right)  - 1 \right) -  2{\pi}i\,\sum_{i=1}^{N}\left( t + i- {\half} \left( N+1 \right) \right) {\s}_{i}}
A more interesting theory is obtained by taking
$$
{\CR} = V\otimes V^{*} \otimes {\CL} \oplus V \otimes {\CF} \oplus V^{*} \otimes  {\tilde\CF} \ .
$$
which corresponds to the theory with the $H^{\rm max} = U(L) \times U(L) \times U(1)$ global symmetry group.
Here $V = {\bC}^{N}$ is the $N$-dimensional fundamental representation of $G$,
${\CF} \approx {\bC}^{L}$, ${\tilde\CF}\approx {\bC}^{L}$ are the $L$-dimensional fundamental representations of the first and the second $U(L)$ factors in the flavour group, and ${\CL}$ is the standard one-dimensional representation of the global group $U(1)$. In simple terms, this theory has the matter content of the four dimensional $N_{c} = N$, $N_{f} = L$,
${\CN}=2$
gauge theory with fundamental hypermultiplets, however, the supersymmetry is half that of the four dimensional theory.
This theory has $2L+1$ twisted mass parameters
(we skip tildes from now on): $({m}_{a}^{\rm f}, {m}_{a}^{\bar{\rm f}} )_{a=1}^{L},  {m}^{\rm adj} = -iu$.
Upon integrating out the matter fields and the $W$-bosons we get the theory of the abelian vector multiplet with the effective twisted superpotential:
\eqn\effspqfq{\eqalign{&
{\tilde W}_{{\tilde Q}{\Phi}Q}^{\rm eff}({\s}) = \cr
& \qquad \sum_{i=1}^{N} \sum_{a=1}^{L} \left[ \left( {\s}_{i} + m_{a}^{\rm f} \right) \left( {\rm log} \left( {\s}_{i} + m_{a}^{\rm f} \right) - 1 \right) +
\left( - {\s}_{i} + {m}_{a}^{\rm \bar f}  \right) \left( {\rm log} \left( - {\s}_{i} + {m}_{a}^{\rm \bar f}  \right) - 1 \right) \right] \cr
& \qquad\qquad\qquad\qquad + \sum_{i,j=1}^{N} \left( {\s}_{i} - {\s}_{j} + m^{\rm adj} \right) \, \left( {\rm log} \left( {\s}_{i} - {\s}_{j} + m^{\rm adj} \right) - 1 \right) \cr
& \qquad\qquad\qquad\qquad\qquad\qquad - 2\pi  i  \sum_{i=1}^{N} \left( t + i  -
{\half} (N+1) \right) {\s}_{i} \cr}}
The generic twisted masses are incompatible with any tree level superpotential. However, for
the special choice of the twisted masses one can turn on the tree level superpotential. Its variation does not change the effective twisted superpotential \effspqfq\ though. We shall discuss this point later.

\subsubsec{Three dimensions}

Consider now the theory on ${\bR}^{2} \times {\bS}^{1}$.  It suffices to make all the fields depend on an extra coordinate  $x^{2} = y$, $y \sim y + 2{\pi}$.
Since the translations in $y$ are the global symmetry of the theory we can turn on the corresponding twisted mass\foot{The space of fields is of course acted on by ${\rm Diff}({\bS}^{1})$, but the Lagrangian is invariant only under ${\bS}^{1}$, the translations.}  ${\tilde m}_{{\bS}^{1}}$. This is equivalent to
promoting the real part of the complex scalar in the vector multiplet to the
covariant derivative:
\eqn\scov{\eqalign{ & {\s}(t, x) \longrightarrow
{1\over R}{\p}_{y} + {\s}(t, x, y) \ , \cr
& {\bar\s} (t, x) \longrightarrow - {1\over R}{\p}_{y} + {\bar\s}(t, x, y) \cr}}
where $R$ is the radius of the circle ${\bS}^{1}$. In other words,
\eqn\sigm{{\s}  =  {1\over R} A_{y} +  {\s}_{\bR}}
where $A_{y}$ is the gauge field component (the $y$ coordinate being dimensionless the $A_{y}$ field is dimensionless too, while ${\s}$ has a dimension of mass).
The twisted mass corresponding to the translations is
${\tilde m} =  {i\over R}$.
Thus, the  Kaluza-Klein modes with momentum $n$, $n \in {\bZ}$, have the corresponding twisted mass
\eqn\twms{{\tilde m}_{n} = {in \over R}}
To compute the effective twisted superpotential, it suffices to enumerate the Kaluza-Klein modes and sum up their contributions. One needs to use a kind of zeta-regularization, which can be
justified, e.g. by  topological field theory methods \nikfive.

{}For definiteness let us consider the contribution of a matter field in the representation $\CR$ of the gauge group. Let
$\tilde\bm$ denote the ordinary two dimensional twisted mass, corresponding to the centralizer of $G$ in $\CR$ which preserves other couplings of the theory, such as the superpotential. We assume $\tilde\bm$ sufficiently generic so that all the modes of the corresponding matter multiplet are massive. The sum over the Kaluza-Klein
modes gives:
\eqn\kksum{\eqalign{{\tilde W}^{\rm eff}_{\rm matter} ({\s}) \, = \, & {\tr}_{\CR} \left[ \,
\sum_{n \in {\bZ}}  \, \left( {\sigma} + {\tilde \bm} + {in\over R} \right) \left( {\rm log}\left( {\sigma} + {\tilde \bm} + {in\over R} \right)  - 1 \right) \right] \sim \cr
& \qquad\qquad
   {1\over 2{\pi}R} {\tr}_{\CR}\, \left[{\rm Li}_{2} \left( e^{-2{\pi}R\left( {\s} + {\tilde \bm} \right)} \right) \right] \cr}}
In addition to the matter-induced twisted superpotential we also have a contribution of the $W$-bosons:
\eqn\kksupw{\eqalign{{\tilde W}^{\rm eff}_{\rm gauge} = &  - {\tr}_{{\bg}/{\bt}}
\, \left[
{1\over 2{\pi}R} {\rm Li}_{2} \left(
e^{-2{\pi}R{\s}} \right) \right] = {{\pi}R\over 2}
{\tr}_{\rm adj} \, \left( {\s}^{2} \right) + 2 {\pi}i\,  \langle {\rho}, {\s} \rangle  \cr}}
where we used:
\eqn\litwo{{\rm Li}_{2} ( e^{-x} ) +
  {\rm Li}_{2} (e^{x} ) = {{\pi}^{2}\over 3} - i {\pi}x - {x^{2}\over 2}}
and dropped an irrelevant constant. The quadratic term in \kksupw\ corresponds to the anomaly-induced Chern-Simons interaction \ref\redlich{A.~Redlich, ``Gauge Noninvariance and Parity Nonconservation of Three-Dimensional Fermions,'' \prl{52}{1984}{18-21}},
\ref\antti{A.J.~Niemi, G.W.~Semenoff, ``Axial Anomaly Induced Fermion Fractionization and Effective Gauge Theory Actions in Odd Dimensional Space-Times'',
\prl{51}{1983}{2077}} in the three dimensional theory.

\subsubsec{Four dimensions}

We can lift the theory to the ${\CN}=1$ supersymmetric Yang-Mills theory (with matter), compactified on a two-torus ${\bT}^{2}$. Again, we can view the lift to four dimensions as the two dimensional theory with the infinite number of fields, which depend on the two additional coordinates $(y,z)$, with $y \sim y + 2\pi$, $z \sim z + 2\pi$. The theory is regularized by the twisted masses corresponding to the translations along ${\bT}^{2}$.
We choose one of the masses to be ${i\over R}$, then the other is
${i \tau \over R}$. Here $\tau$ is the complex modulus of ${\bT}^{2}$.
The normalized holomorphic coordinate on ${\bT}^{2}$ is given by: $w = {1\over 2\pi} \left( y + {\tau}z \right)$. The gauge theory is sensitive to the metric on the torus and a two-form, the so-called $B$-field, via the coupling
\eqn\bfclp{\int_{{\bR}^{2} \times {\bT}^{2}} B \wedge {\tr}F \ .}
Similarly to the  three dimensional lift of the previous section the field $\sigma$ gets promoted to the covariant derivative operator (${\tau}_{2} = {\rm Im}{\tau}$):
\eqn\prom{{\s}(t,x) \rightarrow {{\tau}_{2}\over i {\pi} R} {\bar \p} + {\s}(t,x,y,z), \quad {\bar \s}(t,x) \rightarrow {{\tau}_{2}\over i{\pi}R} {\p} + {\bar\s}(t,x,y,z)}
where
$$
{\pb} = {i\pi \over {\tau}_{2}} \left( {\p}_{z} - {\tau} {\p}_{y} \right)
$$
The
invariance under the large gauge transformations now translates to
the double-periodicity of the twisted superpotential:
\eqn\dblpr{{\s} \to {\s} + {i\over R} \left( m + n {\tau}\right)\, , \,   m,n \in {\bZ}}
The effective twisted superpotential is
given by (${\rm q} = {\exp} \, 2\pi i \tau$):
\eqn\wsupell{\eqalign{&{\tilde W}^{\rm eff}
= {{\pi}R\over 2} {\tr}_{\CR} ( {\s} + {\tilde\bm} )^{2} + {{\pi}i\tau \over 6} {\tr}_{\CR} ( {\s} ) \, + \cr
& \qquad\qquad\qquad\qquad {1\over 2\pi R}\sum_{n=1}^{\infty} {\tr}_{\CR}
\left[ \,
{\rm Li}_{2} \left( {\rm q}^{n-1} e^{-2{\pi}R ( {\s}+ {\tilde\bm})} \right)\, -
{\rm Li}_{2} \left( {\rm q}^{n} e^{2{\pi}R ( {\s} + {\tilde\bm} )} \right) \right]\cr
&  \qquad = \, {{\pi}R\over 2} {\tr}_{\CR} ( {\s} + {\tilde\bm} )^{2} + {{\pi}i\tau \over 6} {\tr}_{\CR} ( {\s} )  +  {1\over 2\pi R} \sum_{n \in {\bZ}_{\neq 0}} {{\tr}_{\CR}
\left[ \,  e^{2\pi R n \left( {\s} + {\tilde\bm} \right)} \right] \over n^{2} ( 1 - {\rm q}^{n} ) }  \cr}}
plus linear terms.

\subsec{Supersymmetric vacua of ${\CN}=2$ theories}

The only local gauge invariant of the abelian gauge field in two dimensions is the field strength $F_{01}$ which
is subject to the only global constraint:
\eqn\qcs{{1\over  2{\pi} i}\int_{\Sigma} F^i = m^i \in {\bZ}}
i.e. the integrality of the magnetic flux. In addition, the global invariants of the ${\bT}$-valued gauge field include the holonomies,
which are irrelevant for our discussion at the moment.

In order to minimize the
potential energy and find the vacua of the theory we promote
$F_{01}^{\a}$ to the independent
fields, while adding at the same time
the term
\eqn\dl{\sum_{i=1}^{r} n_i \int_{\Sigma} F^i}
to the action (cf. \BlThlgt,\btverlinde,\gerasimov). Following \issues,
the shift \dl\  is equivalent to the shift
\eqn\wshft{{\tilde W}^{\rm eff} ({\s}) \longrightarrow
{\tilde W}^{\rm eff}_{\vec n} ({\s})
= {\tilde W}^{\rm eff} ({\s})
- 2\pi i\sum_{i=1}^{r} n_i {\s}^i}
where now $H\pm i F_{01}$ (cf. \sig\ ) are two independent auxiliary fields, which can be integrated out. Thus the target space of the effective sigma model becomes, {\it a priori}, disconnected, with ${\vec n}$ labeling the connected components. In fact, the actual connected components are labeled by the equivalence classes of ${\vec n}$
up to the action of the monodromy group (the effective superpotential
is not a univalent function of ${\s}$). The potential on the component, labelled by $\vec n$ is given by (note that unlike the standard expressions involving ``$\ldots {\rm min}_{n} ( x + 2{\pi}n)^{2}$, which follows from the pair
creation in the background electric field induced by the theta angle$\ldots$ '' it is consistent with supersymmetry and holomorphy):
\eqn\potu{U_{\vec n} = {\half}
{\rm g}^{ij} \left( -2\pi i n_i + {{\p\tilde{W}}^{\rm eff} \over {\p{\s}^i}} \right) \left( + 2\pi i n_j + {{\p\tilde{\bar W}}^{\rm eff} \over {\p{\bar\s}^j}} \right)}
The minima of the effective potential \potu\ are thus the solutions of the equations:
\eqn\vace{
{1\over 2\pi i}{{\p {\tilde W}^{\rm eff}(\sigma)} \over {\p \s}^i} =  n_i}
This equation is derived under very general conditions. Everything is hidden in ${\tilde W}^{\rm eff}$. The $n_{i}$ dependence in \vace\   can be eliminated by exponentiating both sides:
\eqn\mainagain{\mathboxit{
{\exp} \left( {\p {\tilde W}^{\rm eff}}(\sigma) \over {\p \s}^{i}  \right) =  1}}

\subsec{Examples of the vacuum equations}

\subsubsec{Old examples: asymptotically free theories}

Asymptotically free theories are certain limits of asymptotically conformal theories. Since our main examples are asymptotically conformal for completeness we give couple of examples of asymptotically free theories first.

\ndt{$\underline{{\bC\bP}^{L-1}\ \rm model}$.} $G = U(1)$, ${\CR} = R_{+1} \otimes {\bC}^{L}$, where $R_{+1}$ is a one-dimensional charge $+1$ representation of $U(1)$. From twisted effective superpotential of this model
we immediately derive:
\eqn\cpmv{
\prod_{a=1}^{L} ({\s} + {\tilde m}_{a} ) =
q \equiv e^{2\pi i\, t}}
which implies that the model has
$L$ isolated vacua, and the theory at each vacuum is massive, for the
generic values of the twisted masses 
${\tilde m}_{a}$. For vanishing
twisted masses the equation \cpmv\ simplifies to ${\s}^{L} = q$
which is the famous quantum cohomology ring of ${\bC\bP}^{L-1}$. For the generic twisted masses the equation \cpmv\ describes the $U(L)$-equivariant quantum cohomology $H^{*}_{U(L)}({\bC\bP}^{L-1})$ ring.

The next example is that of the

\ndt
{$\underline{{\rm Grassmanian\ } Gr (N, L )\ {\rm model}}$.} $G = U(N)$ and $\CR = {\bC}^{N} \otimes {\bC}^{L}$. Using the effective twisted superpotential of this model we derive ($q = e^{2\pi i t}$):
\eqn\grv{\prod_{a=1}^{L} ({\s}_{i} +
{\tilde m}_{a} ) =  (-1)^{N+1}\, q\, , \ i = 1, \ldots , N}
We should supplement the equations \grv\ with the condition that ${\s}_{l} \neq {\s}_{m}$ for $l \neq m$ and identify the solutions which differ by the permutations of ${\s}_{l}$'s. In other words, the
equations \grv\ should be viewed as equations on the elementary symmetric functions
\eqn\smmfn{c_{l} = \sum_{i_{1} < \ldots < i_{l}} {\s}_{i_{1}} {\s}_{i_{2}}\ldots {\s}_{i_{l}}}
which can be compactly written using the gauge invariant order parameter ${\bQ}(x)$,
\eqn\bxt{{\bQ}(x) \equiv {\det} ( x  - {\s} ) =  \prod_{i=1}^{N} ( x - {\s}_{i} ) = x^{N} + \sum_{i=1}^{N} (-1)^{i}c_{i} x^{N-i} \ , } which we
shall call the Baxter-Chern (BC) order parameter, as:
\eqn\baxtrgr{\prod_{a=1}^{L} ( x +
{\tilde m}_{a} )  + (-1)^{N} q \, = t(x) {\bQ}(x)}
for some polynomial $t(x)$ of degree $L - N$,
$$
t(x) = x^{L-N} + \sum_{j=1}^{L-N} t_{j}x^{L-N-j}\ .
$$
This polynomial is
uniquely fixed in terms of $c_{i}$'s from the equation
\baxtrgr\ by expanding both sides
at $x=\infty$ and equating the
coefficients of $x^{L-N-j}$, $j=1, \ldots , L-N$.
In the classical limit $q \to 0$ the
polynomial ${\bQ}(x)$ is essentially the $U(L)$-equivariant Chern polynomial of the tautological rank $N$ bundle $E$ over  the Grassmanian ${\rm Gr}(N, L)$, while $t(x)$ is the $U(L)$-equivariant Chern polynomial of the tautological dual bundle $E^{\perp}$ of rank $L-N$. The relation \baxtrgr\ then reads simply as the consequence of the exactness of the sequence:
$$
0 \longrightarrow E \longrightarrow F \approx {\bC}^{L}
\longrightarrow E^{\perp} \longrightarrow 0
$$

\ndt
{$\underline{{\CN}=2^{*}\, \rm theory.}$}

The example of the pure ${\CN}=4$ theory broken down to ${\CN}=2$ by the twisted mass term for the adjoint chiral multiplet is the first example where the supersymmetry is broken, for $N>1$.  Here $G=U(N)$, $SU(N)$, $SO(N)$, $Sp(N)$ and ${\CR} = {\bg} \otimes {\bC}$, i.e. the adjoint representation.
Using
\effspadj\ we derive:
\eqn\adjv{\prod_{j=1}^{N} {{\s}_{i} - {\s}_{j} + m \over {\s}_{i} - {\s}_{j} - m} = - q\ , }
which can be neatly rewritten using our ${\bQ}$-operator \bxt\ again:
\eqn\adjvq{{\bQ} (x + m)  + q\,
{\bQ}(x- m) = ( 1 + q ) \, {\bQ}(x)}
It is easy to see that this equation has no solutions for $\sigma_{i}$'s for $N>2$, or for $N =1, q\neq 1$ and has a valley of solutions for $N=1$, $q=1$.

\ndt
{$\underline{\rm Hitchin\ theory.}$}

The model studied in \higgs, \gerasimovshatashvili, \gstwo\ corresponds to the ${\CN}=2^{*}$ theory with the tree level {\it twisted} superpotential\foot{In most of the discussion we have the tree level {\it superpotential}, rather then the tree level
{\it twisted superpotential} turned on.}:
\eqn\lw{{\tilde W}( {\s} ) = {{\l} \over 2}  {\tr}\, {\s}^2 \ ,}
which corresponds to the two-observable representing the K\"ahler form on the Hitchin's moduli space ${\CM}_{H}$. This leads to the change in the right hand side of \adjv:
\eqn\adjvv{\prod_{j=1}^{N} {{\s}_{i} - {\s}_{j} + m \over {\s}_{i} - {\s}_{j} - m} = {\exp} \, 2\pi i {\l} \, {\s}_{i} \ , }
and one now gets solutions for
${\s}_{i}$'s for all $N$.  The topological twist of this theory, introduced in \higgs\ and was studied in detail in \gerasimovshatashvili, \gstwo.

\subsubsec{New examples: asymptotically conformal theories}

Our main example will be the $U(N)$ gauge theory with $L$ fundamental chiral multiplets ${\bQ}_{a}$, $L$ anti-fundamental chiral multiplets ${\tilde\bQ}^{a}$, and one adjoint chiral multiplet ${\bf \Phi}$. This matter content corresponds to the gauge theory with extended supersymmetry, ${\CN}=4$, which is the dimensional reduction of the
four dimensional ${\CN}=2$ theory. The adjoint
$\bf\Phi$ is a part of the vector multiplet in four dimensions, while the chiral fundamental and anti-fundamentals combine into the four dimensional hypermultiplet in the fundamental representation. We are dealing, therefore, with the matter content
of the four dimensional ${\CN}=2$ theory with $N_{c} = N$, $N_{f} = L$. If the superpotential $
\sum_{a} {\tilde Q}^{a}{\Phi}Q_{a}$ is added, then the theory {\it does} have the four dimensional ${\CN}=2$ supersymmetry.

Since the gauge group has a center
$U(1)$ one can turn on the Fayet-Illiopoulos term, and the theta angle as we already explained in {\bf Section 2}, which we combine into a
complexified coupling ${\vt} \mapsto t = {{\vt} \over {2\pi}} +  i r$.

First, we consider the theory with general twisted masses for the chiral fundamentals, anti-fundamentals, and the adjoint field (which is compatible only with the zero superpotential). We then turn on the superpotential and discuss the consequences.
\bigskip
\bigskip
\ndt {$\underline{\rm Two\ dimensions}$.}

Using \vace\ with \effspqfq\ we arrive at the
equations for vacua (we shift $t$ by  $L/2$ to avoid extra phases in the right hand side):
\eqn\vacei{\prod_{a=1}^{L} {{\s}_i + m_{a}^{\rm f} \over {\s}_i - m_{a}^{\bar{\rm f}}} =  - e^{2\pi i {t}}\prod_{j =1}^{N}
{{\s}_i - {\s}_j - m^{\rm adj} \over
{\s}_i - {\s}_j + m^{\rm adj}}}
The equation \vacei\ is written in terms of the eigenvalues ${\s}_{i}$ of the complex scalar $\s$. The equations have solutions related by permuting ${\s}_{i}$'s. These solutions are physically equivalent. It is better to formulate \vacei\ directly in the gauge invariant terms. This is done, similar to Grassmanian case above,  with the help of the BC order parameter \bxt.
The equation \vacei\ is equivalent to:
\eqn\baxq{a(x)\,  {\bQ}(x+m^{\rm adj} )\,
+ \, e^{2\pi i t}\, d (x) {\bQ} (x - m^{\rm adj} )\,  = \, t(x) {\bQ}(x)}
where:
\eqn\aadt{a(x) =  \prod_{a=1}^{L} ( x + m_{a}^{\rm f} ) \ ,
\ d(x) = \prod_{a=1}^{L} ( x - m_{a}^{\rm \bar f})}
and $t(x)$ is an unknown polynomial of degree $L$.

\ndt {$\underline{\rm Three\ dimensions}$.} If we take the analogous theory in three dimensions, compactified on a radius $R$ circle, the resulting vacuum equations would look like:
\eqn\vaceii{\prod_{a=1}^{L} {{\rm sinh} \left( {\pi}R \left( {\s}_i + m_{a}^{\rm f} \right) \right) \over {\rm sinh} \left( {\pi}R \left( {\s}_i - m_{a}^{\bar{\rm f}} \right) \right) } = -e^{2\pi i t}\prod_{j =1}^{N}
{{\rm sinh} \left( {\pi}R \left( {\s}_i - {\s}_j - m^{\rm adj} \right) \right) \over {\rm sinh} \left( {\pi}R \left(
{\s}_i - {\s}_j + m^{\rm adj}\right)\right)}}
Notice the invariance of the eqs. \vaceii\ under the transformations:
\eqn\ggtrs{{\s}_i \longrightarrow
{\s}_i + { in_i \over R}\, , \, n_i \in {\bZ}}
and the permutations of ${\s}_i$'s. This invariance is the  affine Weyl group symmetry, the residual gauge invariance, whose origin is the gauge transformations of the form: $$g(y) = {\rm diag}\,
\left( \, e^{in_{1} y}, \ldots , e^{in_{N}y} \, \right)\ .$$
The equations \vaceii\ can be also analyzed in the gauge invariant fashion using the BC operator. The order parameters of the three dimensional theory compactified on the circle ${\bS}^{1}$ are contained in the trigonometric polynomial (cf. \nikfive):
\eqn\qtrig{{\bQ} (x) = 2^{N}e^{{\pi}R N {\hat y}}\prod_{i=1}^{N}
{\rm sinh} \left( {\pi}R \left( {\hat y} - {\s}_{i} \right)\right) =
 x^{N} + u_{1} x^{N-1} + \ldots + u_{N} \, , }
where
\eqn\xye{ x = {\exp} \, ( 2 \pi R {\hat y} )}
The equations \vaceii\ are equivalent to the difference equation:
\eqn\baxeii{a(x) {\bQ} ( x {\hat q} ) + q\, d(x) {\bQ} ( x {\hat q}^{-1} ) = t(x) {\bQ}(x)}
where $q = e^{2\pi i t}$,
\eqn\adtx{\eqalign{& \quad\qquad {\hat q} = e^{2{\pi}R m^{\rm adj}}\, , \cr
& a(x) = \prod_{a=1}^{L} ( x e^{\pi R m_{a}^{\rm f}} - e^{-\pi R m_{a}^{\rm f}} )\, ,  \cr
& d(x) =  \prod_{a=1}^{L} ( x e^{-\pi R m_{a}^{\rm \bar f}} - e^{+\pi R m_{a}^{\rm \bar f}} ) \cr}}
and $t(x)$ is a polynomial to be determined.

In the limit $R \to 0$ with all other parameters kept finite we recover the two dimensional story.

\ndt {$\underline{\rm Four\ dimensions}$.} The four dimensional gauge theory
with the similar field content, compactified on a two-torus with
the modular parameter $\tau$,
will lead to the elliptic generalization of \vaceii:
\eqn\vaceiii{\prod_{a=1}^{L} {{\Theta}_{1} \left( {\pi}R \left( {\s}_i + m_{a}^{\rm f} \right) \right) \over {\Theta}_{1} \left( {\pi}R \left( {\s}_i - m_{a}^{\bar {\rm f}} \right) \right) } = - e^{2\pi it}\prod_{j =1}^{N}
{{\Theta}_{1} \left( {\pi}R \left( {\s}_i - {\s}_j - m^{\rm adj} \right) \right) \over {\Theta}_{1} \left( {\pi}R \left(
{\s}_i - {\s}_j + m^{\rm adj}\right)\right)}}
where (in this section $\rm q$ denotes
${\exp} (2{\pi}i \tau )$):
\eqn\btfz{{\Theta}_{1}({\xi})  =  - i {\rm q}^{1\over 8}
\left( e^{\xi} - e^{-\xi} \right)
\prod_{m=1}^{\infty}
\left( 1 - {\rm q}^{m} \right)
\left( 1 - {\rm q}^{m} e^{2 \xi}\right)
\left( 1 - {\rm q}^{m} e^{-2 \xi}\right)}
The gauge invariance of the equations \vaceiii\ is more subtle then that of its three and two dimensional counterparts.
We have the gauge transformations of the form:
\eqn\gf{g(y,z) = {\rm diag}
\left( e^{i n_{1}y - i m_{1}z}, \ldots , e^{i n_{N}y - i m_{N}z} \right) \, , \qquad n_{i}, m_{i} \in {\bZ}} which act on ${\s}$ as follows:
\eqn\strf{{\s}_{i} \mapsto {\s}_{i} +
{i\over R} \left( n_{i} + m_{i} {\tau} \right)}
The shifts by $n_{i}$'s are clearly a symmetry of \vaceiii. The shifts by $m_{i}$'s are more subtle.
It turns out that to maintain the invariance of \vaceiii\ under these shifts one has to assume that
$$
\sum_{a} \left( m_{a}^{\rm f} + m_{a}^{\bar {\rm f}} \right)  = -N m^{\rm adj}
$$
and that $t$ transforms under the $U(1)$
subgroup of the $U(N)$ gauge transformations.
The physics of this phenomenon is rather deep, as it involves the chiral anomalies of the charged fermions in four dimensions \bass.

\newsec{Spin chains and Bethe ansatz}

In this section we give a swift review of the integrable spin chains at the example of the $XXX$ spin chain for $SU(2)$. We also briefly
mention other models like $XXZ$, $XYZ$, spin chains with other groups, various
boundary conditions, various limits, such as the one-dimensional Bose gaz, the one-dimensional Hubbard model, etc. The so-called Yang-Yang (YY) function $Y({\l})$ plays the central r\^ole in our discussion. Its critical points are the solutions of Bethe equations. These equations determine the spectrum of integrable hamiltonians. That the equations determining the spectrum have a potential is a highly non-trivial consequence of the rich algebraic structure behind these systems.
It is also the cornerstone of our correspondence with the gauge theories.

\subsec{XXX spin chain}

The Heisenberg spin chain, also known as the $SU(2)$ $XXX$ spin chain, is defined on the one dimensional length $L$ lattice. At each lattice point one has the spin $s={1 \over 2}$ representation of $SU(2)$, and the Hilbert space of the system is the tensor product ${\CH}_{L} =
{\bC}^{2} \otimes {\bC}^{2} \otimes \ldots \otimes {\bC}^{2}
$. The Hamiltonian $H_{\rm Heis}$ acts in ${\CH}_{L}$. It is written in terms of generators ${\vec S}_a = {i \over 2} {\vec\s}_{a}$
where $a$ denotes the position on the lattice of the spin $s={1 \over 2}$ representation of $SU(2)$ and has the nearest-neighbor interaction form:
\eqn\hxxx{H_{\rm Heis} = J\, \sum_{a=1}^{L}(\, S_{a}^{x} S_{a+1}^{x} + S_{a}^{y}S^{y}_{a+1} + S^{z}_{a}
S_{a+1}^{z})}
The boundary conditions are quasi-periodic:
\eqn\qper{
\vec S_{L+1}= e^{{\ihalf} {\vt}{\s}_{3}}\vec S_{1} e^{- {\ihalf} {\vt} {\s}_{3}}\ .}
In other words we identify ${\CH}_{L}$ with the subspace ${\CH}_{L}^{\vt} \subset \left( {\bC}^{2} \right)^{\otimes \infty}$, characterized by \qper.
One can also consider the spin chains defined on an open interval. For the ferromagnet $J>0$  and for the anti-ferromagnet - $J < 0$.

The total spin,
${\vec \bS} = \sum_{a=1}^L {\vec S}_{a}$ commutes with $H_{\rm Heis}$ for ${\vt} = 0$.
The spin projection on the third axis, ${\bS}^{z}$, is a conserved quantity for any ${\vt}$. The corresponding subspace of the Hilbert space, ${\CH}^{N}_{L} \subset {\CH}_{L}$, where ${\bS}^{z} = N - {\half}L$, is sometimes
called the $N$-particle sector.

We study the $N$-particle eigenstates of $H_{\rm Heis}$. The states in ${\CH}^{N}_{L}$
are the linear combinations of the states with
$N$ spins up and $L-N$ spins down. Clearly, the maximal number of spins up or down is
$L$, so $|{\bS}^{z}| \leq {L \over 2}$, and $N \leq L$. The $N$-particle state
$| \Psi \rangle$ can be expanded as:
\eqn\eigenx{| \Psi \rangle = \sum_{1 \leq x_1 < \ldots < x_N \leq L}  {\Psi}(x)\, | x_{1}, \ldots , x_{N}\rangle}
with $| x_{1},\ldots , x_{N}\rangle$ denoting the state in above tensor product with spins up at the
positions $x_{1},\ldots , x_{N}$: $ | x_{1}, \ldots , x_{N}\rangle =
S_{x_{1}}^{+} \ldots S_{x_{N}}^{+}\, {\Omega}$,
where ${\Omega} =
| {\downarrow} {\downarrow} \ldots {\downarrow} \rangle$
is the (pseudo)vacuum, the state with all spins down. It is annihilated by all operators $S_{x}^{-}$, $
S^{-}_{x}\, {\Omega} = 0
$.
The total number of the $N$-particle eigenstates of the Hamiltonian $H_{\rm Heis}$ is $\pmatrix{L \cr N}$, as they can be enumerated by the appropriate functions ${\Psi}(x)$.

\subsubsec{The coordinate Bethe ansatz}

In 1931 H.~Bethe parametrized
\ref\hbethe{H.~Bethe, Z.~Phys. {\it 71}
(1931) 205} these functions by $N$ quasimomentum variables $
p = \left( p_{1}, \ldots , p_{N}\right) \ ,
$
subject to the further equations which we write momentarily. The  ansatz, known as {\it Bethe ansatz}, reads as follows: let
\eqn\an{{\Psi}_{p}(x_1,...x_N)=\sum_{w \in S_N} (-1)^{w} A(p_{w(1)},..,p_{w(N)})
\, {\exp}\, \left( \sum_{j=1}^N ip_{w(j)}x_j \right) \ , }
then the eigenstate of $H_{\rm Heis}$ is given by $|{\bf\Psi}_{p}>=\sum_{1 \leq x_1 < ... < x_N \leq L}  {\Psi}_{p}(x) | x_1,...,x_{N} \rangle$.
The Bethe ansatz expresses the coefficients $A_{p}(x)$  in terms of the two body $S$-matrix $\Sigma(p_1,p_2)$:
$$A(p_{1},...,p_{N})= \prod_{1 \leq j \leq k \leq N} {\Sigma} (p_{j}, p_{k} ), \quad \quad {\Sigma} (p_{j}, p_{k}) =1-2 e^{ip_{k}}+e^{i (p_{j}+p_{k})}
$$
It is more convenient to use the new variables $\lambda_j$  instead of $p_{j}$:
$$e^{ip_{j}}={{\lambda_{j}+{i \over 2}} \over {\lambda_{j}-{i \over 2}}} \ .
$$
In this notation
${\bf\Psi}_{\l}(x)$ of \eigenx\ is an eigenstate of the $H_{\rm Heis}$ if and only if $(\lambda_1,...,\lambda_N)$ satisfy the Bethe equation:
\eqn\baexxx{\left( { {\l}_j + {i \over 2}  \over
 {\l}_j -  {i \over 2}  } \right)^{L} = e^{i\vt}\prod_{k\neq j}
\, { {\l}_j- {\l}_{k} + i   \over  {\l}_j- {\l}_{k} - i  }}
which guarantees that \an\ obeys the twisted boundary conditions \qper.
The energy of the state \eigenx\ is $H_{\rm Heis} {\bf\Psi}_{p} = E_{p} {\bf\Psi}_{p}\ , \, E_{p} = J \left( L - 2 N + 2 \sum_{i=1}^{N}
 {\rm cos} \left( p_{i} \right) \right) \ .$

A similar construction works for an arbitrary spin, when $\vec{\bS}_a$ is in the spin $s_{a}$ representation of $SU(2)$ at every site of a chain.  In addition, the spin sites can be, in some sense, displaced from the symmetric round-the-clock configuration, so that one gets $L$ additional parameters ${\n}_{1}, \ldots , {\n}_{L}$. This model is sometimes called the inhomogeneous $XXX_{s}$ magnet. The corresponding Bethe equations have the form:
\eqn\baa{ \prod_{a=1}^{L}  { {\l}_j - {\n}_{a} + {is_{a}}  \over
 {\l}_j - {\n}_{a} -  {is_{a}}  }  = e^{i\vt}\prod_{k \neq j }
\, { {\l}_j- {\l}_k + i   \over  {\l}_j - {\l}_k - i  }}
The
Hamiltonian for the general local spins is given by a polynomial in the neighbouring spins, which is more complicated  then \hxxx, see \bass\ for details.

\subsubsec{The analytic Bethe Ansatz}

There is yet another interpretation of the Bethe equations \baa, due to \kolyaf, \baxteriii, as the condition for the polynomial function
\eqn\qbax{{\bQ} ({\l}) = \prod_{i=1}^{N} ( {\l} - {\l}_{i} )}
to solve Baxter's equation
\eqn\baxead{ a({\l}) {\bQ} ( {\l} + i )
+ e^{i \vt} d ({\l}) {\bQ} ({\l} - i  ) = t({\l})\, {\bQ}({\l})}
 with the given polynomials:
\eqn\adginh{a({\l}) = \prod_{a=1}^{L}
( {\l} - {\n}_{a} - i s_{a}  ) \, ,
\ \ d({\l}) = \prod_{a=1}^{L}
( {\l} - {\n}_{a} + i s_{a} )}
and some unknown degree $L$ polynomial $t({\l})$.

Indeed, let us define $t({\l})$ as the ratio of the left hand side of \baxead\ and ${\bQ}({\l})$. The absence of poles of $t({\l})$ at the zeroes of ${\bQ}({\l})$, i.e. at ${\l} = {\l}_{j}$, $j=1, \ldots , N$ is equivalent to \baa.

The polynomial $t({\l})$ gives the eigenvalues of the
twisted
transfer matrix
\eqn\twtm{{\sl T}_{\vt}({\l}) =
{\sl A}({\l})
+ e^{i\vt} {\sl D} ({\l})}
which is a central object in the {\it algebraic Bethe Ansatz}  
\ref\stf{L.D.~Faddeev, E.~Sklyanin, L.~Takhtajan,  ``Quantum inverse problem method'', Theor. Math. Phys. {\bf 40:2} (1980) 688-706, \
Teor.Mat.Fiz.40:194-220,1979 (in Russian)},\ref\algbans{L.D.~Faddeev and L.~Takhtajan, 
Russ. Math. Survey {\it 34} (1979) 11},\ref\algbas{L.D.~Faddeev and L.~Takhtajan, 
J.~Sov.~Math {\it 19} (1982) 1596},\ref\fadalg{L.D.~Faddeev, ``Algebraic aspects of Bethe Ansatz'', Int.~J.~Mod.~Phys. {\bf A}10 (1995)
1845-1878, hep-th/9404013},\ref\fadba{L.D.~Faddeev, ``How algebraic Bethe ansatz works for
integrable
model'',  hep-th/9605187}, where it is a trace of monodromy matrix, see \bass\ for details. The quasiclassical limit of the equation \baxead\ defines an analytic curve, whose geometry can be effectively used to write formulae for the matrix elements of local operators \ref\smirnov{F.A.~Smirnov, ``Quasiclassical study of formfactors in finite volume,'' arXiv:hep-th/9802132\semi
``Structure of Matrix Elements in Quantum Toda Chain,'' arXiv:math-ph/9805011}.
    
\subsubsec{Yang-Yang function}

The highly surprising property of the equations \baexxx,\baa\ is that they have a potential \yangyang. If we rewrite \baa\ as $e^{2\pi i\, {\varpi}_{j} ({\l})} = 1$, then the following one-form:
\eqn\kfmr{{\varpi} = \sum_{j=1}^{N} {\varpi}_{j} ({\l}) {\rm d}{\l}_{j} }
is closed, $d{\varpi} = 0$ and ${\varpi} = dY$
\eqn\yang{\eqalign{& Y ({\l}) = \sum_{a=1}^{L} \,{s_{a}\over \pi} \sum_{j=1}^{N} {\hat x}\left({{\l}_{j} - {\n}_{a} \over  s_{a}} \right) + {1\over \pi} \sum_{j,k=1}^{N}
{\hat x} ({\l}_{j} - {\l}_{k} ) +
\sum_{j=1}^{N} {\l}_{j} \left( n_{j} + {\vt\over 2\pi} \right)\cr
& \qquad\qquad \cr}}
where the integers
$n_{j}$ label various branches of the logarithms, and the function ${\hat x}({\l})$ is given by:
\eqn\htth{
{\hat x} ({\l}) = {\l}\, {\rm arctan} \left( {1\over {\l}} \right) + {1\over 2}\, {\rm log} \left( 1 + {\l}^{2} \right) .}

\subsubsec{Higher rank spin groups}

Now imagine the spin operators $\vec{\bS}_{a}$ are realized as the generators of
some simple Lie algebra ${\bk} = {\rm Lie}K$.
Let $r = {\rm rank}({\bk})$. The number of spin sites $L$ and
the excitation level $N$ of our previous models generalize to the vectors:
${\vec L} = \left( L_{1}, L_{2}, \ldots , L_{{\bf r}} \right) , {\vec N}
= \left( N_{1}, N_{2}, \ldots , N_{{\bf r}} \right)$. The twist parameter becomes the ${\bf r}$-tuple of angles:
$\left( {\vt}_{1}, \ldots , {\vt}_{{\bf r}} \right)$, which define an element of the maximal torus of $K$. The Bethe equations
read as follows:
\eqn\baegen{\prod_{a=1}^{L_{\bi}} {{\l}_{i}^{({\bi})} - {\n}_{a}^{({\bi})} + i s_{a}^{({\bi})} 
\over
{\l}_{i}^{({\bi})} - {\n}_{a}^{({\bi})} - i s_{a}^{({\bi})}  } =  e^{i {\vt}_{\bi}}\prod_{{\bj}=1}^{{\bf r}}\
\prod_{j:\, (i,{\bi}) \neq (j,{\bj})}\
{{{\l}_{i}^{({\bi})} - {\l}_{j}^{({\bj})} + {\ihalf} {\CC}_{{\bi}{\bj}}}\over {{\l}_{i}^{({\bi})} - {\l}_{j}^{({\bj})} - {\ihalf}{\CC}_{{\bi}{\bj}}}}}
where the unknowns (Bethe roots) are ${\l}_{i}^{({\bi})}$, ${\bi} = 1, \ldots, {\bf r}$,
$i = 1, \ldots , N_{\bi}$. The equations \baegen\ describe the spectrum of the transfer matrix
acting in the space
$$
{\CH}_{\vec L} = \bigotimes_{{\bi}=1}^{{\bf r}} \otimes_{a=1}^{L_{\bi}}\,
{\CW}_{s^{({\bi})}_{a}}^{({\bi})} \left( {\n}_{a}^{({\bi})} \right)
$$
where ${\CW}_{s}^{(i)}({\n})$, $2s \in {\bZ}_{\geq 0}, {\n} \in {\bC}$ are the so-called
Kirillov-Reshetikhin modules \baekr, the special evaluation representations of the
Yangian ${\CY}({\bk})$ of $\bk$. The matrix ${\CC}_{\bi\bj}$ in \baegen\ is the
Cartan matrix of $\bk$.

The equations \baegen\ also have a YY function,
see \bass\ for details.
The most general closed spin chains correspond to yet more general representations of the Yangian $Y({\bk})$, not necessarily the Kirillov-Reshetikhin ones. These representations ${\CW}_{\vec\bP}$ are characterized by the highest weights, which are given by an $r$-tuple ${\vec\bP}$ of monic polynomials, called Drinfeld polynomials:
\eqn\drpolyi{{\vec\bP} = \left( P_{1}({\l}), P_{2}({\l}), \ldots , P_{{\bf r}} ({\l}) \right)}
For example, in the case of ${\bk} = {\bs\bl}_{2}$, the inhomogeneous spin chains were characterized by the polynomials $a({\l})$ and $d({\l})$. These polynomials enter Baxter's equations \baxead. These two polynomials can be related to the single Drinfeld polynomial $P_{1}({\l})$, as it should be, since the rank of ${\bf sl}_{2}$ is equal to one:
\eqn\drinfad{{a({\l}) \over d({\l})} = {P_{1}({\l} + {\ihalf}) \over P_{1}({\l}- {\ihalf})}}
Explicitly (${\hat s}_{a} = s_{a} - {\half}$):
\eqn\drrkone{P_{1}({\l}) = \prod_{a=1}^{L}\, \prod_{m_a=- {\hat s}_a}^{{\hat s}_{a}} \, \left( {\l} - {\n}_{a} + i m_{a} \right)}
In the general case the Bethe roots again
form $\bf r$ groups $\left( {\l}_{i}^{({\bi})}\right) $, ${\bi} = 1, \ldots , {\bf r}$, $i = 1, \ldots , N_{\bi}$.
The general Bethe equations can be written for the simply-laced $\bk$, for each $({\bi}, i)$, as:
\eqn\baebksl{{P_{\bi} ({\l}^{({\bi})}_{i} + {\ihalf} ) \over
P_{\bi} ({\l}^{({\bi})}_{i} - {\ihalf} )} =
e^{i {\vt}_{\bi}} \prod_{{\bj}=1}^{{\bf r}} \prod_{j=1}^{N_{\bj}}
{{\l}^{({\bi})}_{i} - {\l}^{({\bj})}_{j} + {\ihalf}  {\CC}_{\bi\bj} \over
{\l}^{({\bi})}_{i} - {\l}^{({\bj})}_{j} - {\ihalf}  {\CC}_{\bi\bj} }\ }
There exists also the generalizations to the non-simply laced $\bk$, and some partial results
for the affine case as well, see \bass\ for details
and references.

The equations \baebksl\ can be also written in the form of Baxter-like equations for ${\bf r}$
polynomial functions ${\bQ}_{\bi}({\l}) = \prod_{i=1}^{N_{\bi}} ( {\l} - {\l}^{({\bi})}_{i} )$, either directly using \baebksl, see \bass, or using the theory of $q$-characters \ref\fr{E.~Frenkel, N.~Reshetikhin, ``The q-characters of representations of quantum affine algebras and deformations of W-algebras,'' 	arXiv:math/9810055v5}, or, for ${\bk} = su({\bf r}+1)$, using the discrete Hirota equations \ref\krichwieg{I.~Krichever, O.~Lipan, P.~Wiegmann, A.~Zabrodin, ``Quantum Integrable Systems and Elliptic Solutions of Classical Discrete Nonlinear Equations'', arXiv:hep-th/9604080}.

\subsec{Anisotropic chains}

The model with the \hxxx\ Hamiltonian can be
generalized to the anisotropic situations:
\eqn\hxyz{H_{\rm Heis} = \, \sum_{a=1}^{L}(\, J_{x} S_{a}^{x} S_{a+1}^{x} + J_{y} S_{a}^{y}S^{y}_{a+1} + J_{z}S^{z}_{a}
S_{a+1}^{z})}
with the general anisotropy parameters $J_{x}, J_{y}, J_{z}$. These more general spin chains
(the $XXZ$, $XYZ$, or the $8$-vertex model \baxter)
also admit the Bethe ansatz, with the Bethe equations \baa\ replaced by the trigonometric or elliptic analogues.

\newsec{The Dictionary}

In this section we present the explicit bridge between the two topics of our story, the dictionary, relating the quantum integrable spin chains and the ${\CN}=(2,2)$ supersymmetric gauge theories in two dimensions.

We do it here at the example of the inhomogeneous twisted $XXX_{s}$ spin chain and
a certain $U(N)$ gauge theory in two dimensions. This map extends to other examples presented above and more, see \bass\ for details.

The foundation of our dictionary is of course the observation that the vacuum equation for the gauge theory \main\ coincides with Bethe equation in the integrable theory (which we formulate in some generality in \baa, \baebksl):

\centerline{\it The effective twisted superpotential corresponds to the YY function}

Actually, the entries of the YY function
are dimensionless, while the vacuum equation \main\ is written for ${\s}$, which has the dimension of mass. The precise relation reads as
follows:
\eqn\yytwsp{\mathboxit{\eqalign{ u \, Y ({\l} ; s, {\n} )
\, & \, =  \, {\tilde W}^{\rm eff} ({\s}; s, {\m} ) \cr
  {\l}_{i}\, u \, &\, = {\s}_{i} \cr
  {\n}_{a}\, u \, &\, = {\m}_{a} \cr}}}
where $u$ is the particular twisted mass, corresponding to the $U(1)$ symmetry breaking the ${\CN}=4$ supersymmetry of the theory we
present below, down to ${\CN}=2$. 

Of course this is only a starting point leading to precise identification of two theories -- the vacuum structure, including the vacuum expectation values of the (twisted) chiral operators on the gauge theory side and the entire spectrum of all integrable Hamiltonians on the spin chain side. The Baxter operator(s)
${\bQ}_{\bi}({\l})$ are identified, up to the rescaling ${\l} \to x = {\l} u $, $Q_{\bi} ({\l}) \to u ^{-N_{\bi}} Q_{\bi} (x)$, with the BC order parameters of the gauge theory.

\subsec{The ${\tilde Q}{\Phi}Q$ theory vs the $XXX_{s}$ spin chain}

Our announced duality maps the inhomogeneous $XXX_{s}$ spin chain to the $U(N)$ gauge theory  with the following matter fields and twisted masses:
\eqn\qfqth{\mathboxit{\matrix{
{\rm Gauge\,  representation} & {\rm Matter \, multiplets} & {\rm Twisted\, mass} \cr
&   \cr
& \cr
{\rm adjoint} &  \qquad \Phi  \qquad & {m}^{\rm adj}  =  - iu \cr
 & \cr
{\bN}  &  \qquad Q_{a} \qquad &
{m}^{\rm f}_{a}  =  -{\m}_{a} + i s_{a} u \cr
&  \cr
{\bar\bN} &  \qquad {\tilde Q}^{a} \qquad & {m}^{\bar{\rm f}}_{a}  =  + {\m}_{a} + i s_{a} u\cr
& \cr
&  a = 1, \ldots , L & \cr}}}
In the absence of superpotential all the parameters are complex numbers, ${\m}_{a}, s_{a}, u \in {\bC}$.
The generic superpotential \wqfq\ breaks the global symmetry group
$U(L) \times U(L) \times U(1)$ down to the subgroup $U(1)$ of the transformations $Q_{a}\mapsto e^{i {\m}_{a}} Q_{a}$, ${\tilde Q}^{a}\mapsto e^{-i {\m}_{a}} {\tilde Q}^{a}$. However, if the matrix-valued function $m_{a}^{b}({\Phi})$ is chosen in a special way, the unbroken subgroup gets enhanced. In particular, when
\eqn\massm{m_{a}^{b}({\Phi}) = {\d}_{a}^{b} {\varpi}_{a}\, {\Phi}^{2s_{a}}\, , \ W_{\tilde Q {\Phi}Q} =
\sum_{a=1}^{L} {\varpi}_{a} \, {\tilde Q}^{a}{\Phi}^{2s_{a}} Q_{a} , }
for some complex constants ${\varpi}_{a}$,
we have the group $U(1)^{L} \times U(1)$ of the transformations
of the form:
\eqn\unbrk{Q_{a} \mapsto e^{i{\m}_{a} - i s_{a}u}Q_{a} \,
, \, {\tilde Q}^{a} \mapsto e^{-i{\m}_{a} - is_{a}u}{\tilde Q}^{a}\, , {\Phi} \mapsto e^{iu} {\Phi}}
In this case we turn on both the superpotential \massm\ and
the twisted masses \qfqth. In order for the superpotential \wqfq\ be a polynomial, we need $2s_{a}$'s be the non-negative integers. Note that the massless ${\CN}=2$, $d=4$
theory has a superpotential 
$W_{0} =\sum_{a=1}^{L} {\tilde Q}^{a}{\Phi} Q_{a}$
which corresponds to $s_{a} = {\half}$.

{}A few comments about the superpotential
\massm\ are in order. In two dimensions the corresponding theory is renormalizable for all half-integer values of $s$. In three dimensions only for $s={1 \over 2}$ or $s=1$ we get renormalizable theory, and in four dimensions - only for $s={1 \over 2}$. One has several approaches to the three and four dimensional theories for the values of $s$ when the superpotentials ${\tilde Q}{\Phi}^{2s}Q$ are not renormalizable: $1.\rangle$  Think about these theories as effective theories arising from a renormalizable  fundamental theory after integrating out some massive modes; $2.\rangle$  View them as the theories with cutoff;  $3.\rangle$ Embed them into string theory, or $4.\rangle$ Abandon them for such values of $s$ altogether. Obviously we do not like to pursue the last option. We describe the details of  $1.\rangle$ in \bass.
\bigskip

\boxit{\bigskip
\ndt {\it Thus, the ${\tilde Q}{\Phi}Q$ theory with 
\eqn\ncnf{N_{c} = N, \ N_{f} = L}
the superpotential \massm\ and the twisted masses
\qfqth\ with the half-integers $s_{a}$ is mapped
to the $N$-particle sector of the twisted inhomogeneous $SU(2)$ $XXX_{s}$ spin chain.
The supersymmetric vacua correspond to Bethe states. The twisted masses correspond to the inhomogeneities ${\n}_{a}$ and the local spins $s_{a}$: 
\eqn\twms{
m_{a}^{\rm f } =  ( - {\n}_{a} - i s_{a} ) u, \ 
m_{a}^{\bar{\rm f}} = ( {\n}_{a} - is_{a} ) u,\ 
m^{\rm adj} =  -iu }
Since the gauge group $U(N)$ has a center, one has an additional parameter, the
complexified theta angle, which is the sum of the theta angle and the Fayet-Illiopoulos term. This parameter is mapped to the twist parameter of the (complexified) spin chain:
\eqn\spch{ t = {1\over 2\pi}{\vt} + i r \, \longrightarrow \, {\vec \bS}_{a+L}  =
e^{-\pi i  t {\s}_{3}} {\vec \bS}_{a}  e^{\pi i  t {\s}_{3}}\ }}}
Note that the only r\^ole of the superpotential $W$  \massm\ is to impose the integrality condition on the $s_{a}$ parameters of the twisted masses \qfqth. It is conceivable that in the absence of $W$ the theory with complex $s_{a}$'s maps to the
${\bf sl}_{2}$ spin chain with possibly infinite dimensional spin representations (still in the $N$-particle sector).

\subsec{Order parameters, Hamiltonians, local operators}

Let us discuss the r\^ole of the BC order parameter and Baxter's equation in the gauge theory.
Define the gauge theory observable, which we shall call the $T$-operator (cf. \baxq):
\eqn\top{{\bT}(x) = a (x) {{\bQ} ( x + m^{\rm adj} ) \over
{\bQ} (x) } + e^{2\pi i t}\, d(x)   {{\bQ} ( x - m^{\rm adj} ) \over
{\bQ} (x) }}
with $a(x), d(x)$ from \aadt. The $T$-operator
is an infinite expansion in $x$, whose coefficients are the gauge invariant functions of $\s$. In a sense, we can view $t(x)$ as the generating function of the twisted chiral ring operators.
Now, the twisted chiral ring is a commutative associative ring \bass\ generated by the coefficients of ${\bQ}(x)$, and the relations which can be concisely formulated as:
\eqn\txrel{{\bT}(x)_{-} \equiv \sum_{n=1}^{\infty} T_{n} x^{-n}  = \{ {\CQ} , \ldots \} }
where ${\CQ}$ is one of the supercharges of the theory.
In other words, in the twisted chiral ring the following equations hold:
\eqn\twchr{T_{n} = 0 \, , \quad n = 1,2, \ldots}
It would be nice to derive
this  from some Ward identities, analogous to the generalized Konishi anomaly
\ref\dsw{F.~Cachazo, M.~Douglas, N.~Seiberg, E.~Witten, ``Chiral Rings and Anomalies in Supersymmetric Gauge Theory'', arXiv:hep-th/0211170,
JHEP {\bf 0212} (2002) {071} }.
In the spin chain the positive coefficients of the expansion of $t(x)$ correspond to the integrable Hamiltonians $H_{k}$ of the model:
\eqn\txpos{{\bT}(x)_{+} = (1 + e^{2\pi i t}) x^{L} + \sum_{k=1}^{L} H_{k} x^{k-1} }
Finally, the gauge theory has non-local operators,
creating soliton states, interpolating between different vacua of the theory. It is natural to identify those with local operators in the spin chain, such as the operator of the local spin ${\vec \bS}_{a}$. The matrix elements of these
operators between the Bethe states, the form-factors \ref\formf{H.~Boos, M.~Jimbo, T.~Miwa, F.~Smirnov, Y.~Takeyama, ``Algebraic representation of correlation functions in integrable spin chains,''arXiv:hep-th/0601132, Annales Henri Poincare {\bf 7} (2006) 1395-1428 }, are worth investigating on the gauge theory side.

\subsec{More general systems}

It is now clear how to generalize this correspondence to other spin systems. Take, for example, the $XXX$ spin chain with the spin group $K$. The Bethe equations \baegen, \baebksl\ tell us what ${\CN}=2$ supersymmetric gauge theory should be taken in order for its vacua represent the Bethe vectors of the spin chain. It is the quiver gauge theory, with the product gauge group $G = U(N_{1}) \times \ldots \times U(N_{{\bf r}})$, the adjoint, bi-fundamental, fundamental, and anti-fundamental matter multiplets, which can be easily read off the
Dynkin diagram of $K$. Again, one turns on the
twisted masses for these various fields, and the
integrality of some of these masses, which in \baebksl\ are represented by the spins $s_{a}^{({\bi})}$ or the components of the Cartan matrix ${\CC}_{\bi\bj}$, comes from the invariance
of the tree level superpotential. 

However, nothing prevents us from turning off the
tree level superpotential. In this way all bets are off, the matrix ${\CC}_{\bi\bj}$ of twisted masses is no longer restricted to be a Cartan matrix, and 
the spins $s_{a}^{({\bi})}$ are no longer restricted 
by any integral considerations. 

If we believe that the rich algebraic structure of the spin chain survives the translation to the
gauge theory then the Yangian ${\CY}({\bk})$
is to be replaced by another algebra, which is worth investigating further, see \bass\ for details.

\newsec{Lifts to higher levels and higher dimensions}

Our two dimensional theories can be lifted to three and four dimensions while keeping the same amount of supersymmetry. The three dimensional theory compactified on a circle would map to the $XXZ$ spin chain (cf. \vaceii\ with \qfqth), the four dimensional theory compactified on ${\bT}^{2}$ (cf. \vaceiii\ ) maps to the $8$-vertex model and the $XYZ$ spin chain.

\subsec{Beyond the known systems}
 
 The correspondence with the supersymmetric gauge theories opens new doors both for the
 quantum integrable systems and for the gauge theories. We already mentioned a possibility of relaxing the integrality of the Cartan matrix ${\CC}_{\bi\bj}$. As another example, we can study other
 four-dimensional constructions leading to an interesting deformation of the would-be-Bethe equations, i.e. the vacuum equations of the compactified four-dimensional supersymmetric gauge theory. 
 
{}We start with the ${\CN}=2$ supersymmetric gauge theory in four
dimensions and compactify it on a two-dimensional sphere
${\bS}^{2}$. Of course,
this compactification breaks supersymmetry,
so we shall have to make a partial twist along ${\bS}^{2}$ to preserve
some fraction of the supersymmetry.

{}This theory is interesting as its low-energy two dimensional dynamics is sensitive to the effects of the four dimensional instantons. The equations \vace\ then contain the complexified four-dimensional coupling
\eqn\fdcplng{{\CT} = {{\theta}\over 2\pi} + {4\pi i \over e^{2}}}
and, for the appropriate four dimensional theory, are modular.

The partial twist is done as follows (cf. \ref\bjsv{M.~Bershadsky, A.~Johansen, V.~Sadov, C.~Vafa, ``Topological Reduction of 4D SYM to 2D $\sigma$--Models'', arXiv:hep-th/9501096, \np{448}{1995}{166-186}}). The holonomy group
of the product manifold ${\Sigma} \times {\bS}^{2}$ with the product metric is $SO(2)_{\Sigma} \times SO(2)_{{\bS}^{2}}$. Here ${\Sigma}$ is the worldsheet of the effective
two dimensional theory. In addition, the
${\CN}=2$ theory has an $SU(2)$ $R$-symmetry group (it can be larger for the theories with matter). The supercharges of the ${\CN}=2$ theory, eight of them, transform as $\left( \pm \half, \pm \half , {\bf 2} \right)$ under $SO(2)_{\Sigma} \times SO(2)_{{\bS}^{2}} \times SU(2)$. Since the two-sphere has no covariantly constant spinors, none of these supercharges are conserved, if the $R$-symmetry group is to be preserved. Now
imagine $SO(2)_{{\bS}^{2}}$ is allowed to act on the $R$-symmetry index. In other words, let us embed
$SO(2)_{{\bS}^{2}} \to SU(2)$, via
\eqn\twst{e^{i {\a}} \mapsto e^{i q {\a} {\s}_{3}}, \ 2q \in {\IZ}}
The eight supercharges now transform as:
$\left( \pm {\half} , \pm \half \pm q \right)$ under $SO(2)_{\Sigma} \times SO(2)_{{\bS}^{2}}$. We now can choose $q = \pm \half$, to make four supercharges have vanishing charge under $SO(2)_{{\bS}^{2}}$. The other four supercharges transform as:
$\left( \pm \half, \pm 1 \right)$ and are not conserved on the two-sphere ${\bS}^{2}$.

\subsubsec{Pure ${\CN}=2$ theory}

As a warmup, consider the compactification of the pure ${\CN}=2$ super-Yang-Mills theory on
${\bS}^{2}$ with the $q = \pm \half$ twist.

The result is the two dimensional theory, with the ${\CN}=2$ supersymmetry in two dimensions. The field content of that theory
contains a massless vector multiplet and a Kaluza-Klein tower
of massive vector and chiral multiplets, all transforming in the adjoint representation of the gauge group.
The lowest massive level comes from the Laplacian eigenstates in the space of the one-forms on
${\bS}^{2}$.

Now we wish to calculate the effective twisted superpotential of the two dimensional theory. We shall take the size of ${\bS}^{2}$ to zero. In this way the massive states  become infinitely massive and ought to decouple.

Now let us turn on the magnetic flux on the two-sphere. More precisely, we can turn on the flux,
for $G = U(N)$,
$$
{1\over 2\pi i}\int_{{\bS}^{2}} F \sim {\rm diag}\left( {\bm}_{1}, \ldots , {\bm}_{N} \right) \ , \, {\bm}_{i} \in {\IZ}
$$
in the maximal torus of the gauge group, determined by the vacuum expectation value of the adjoint Higgs field.
In the presence of the magnetic flux, some of the charged Kaluza-Klein modes become massless and contribute to the effective twisted superpotential. As a result,
the twisted superpotential can be expressed in terms of the prepotential of the four dimensional theory as follows:
\eqn\twstsup{W (a) = \sum_{i=1}^{r}{\bm}_{i}
{{\p\CF}\over {\p}a^{i}}}
where $r = N$ for $G= U(N)$, $r = N-1$ for $G = SU(N)$ (in the latter case there is one more subtlety related to the possibility to turn on the discrete magnetic flux $w_{2} \in {\IZ}_{N}$).
In addition, the unfolding of the two dimensional field strength can be accomplished, as in \potu, by introducing the integral vector
$( {\bn}_{1}, \ldots , {\bn}_{r} )$,
which can be identified with the vector or electric fluxes through the
two-sphere. The twisted superpotential becomes \issues:
\eqn\wsupmn{W (a) = \sum_{i=1}^{r} \left( {\bm}_{i} {{\p\CF}\over {\p}a^{i}} + {\bn}_{i} a^{i} \right) = \oint_{C_{\bm, \bn}} {\l}}
where $\l = p {\rm d}z$ is the Seiberg-Witten differential, and $C_{\bm, \bn} \in H_{1}({\CC}, {\IZ})$ is a cycle on the Seiberg-Witten curve $\CC$,
\eqn\swcrv{{\Lambda}^{N} \left( e^{p} + e^{-p} \right) = z^{N} + u_{1} z^{N-1} + \ldots + u_{N} \ , }corresponding to the charges $({\bm}, {\bn})$.

\subsubsec{The ${\CN}=2^{*}$ theory}

Now, to make things interesting let us add some matter fields.
One of the most beautiful gauge theories in four dimensions is the so-called ${\CN}=2^{*}$ theory. This is the ${\CN}=2$ theory with massive adjoint hypermultiplet. In the ultraviolet this is the ${\CN}=4$ theory, which exhibits $S$-duality.
In the infrared this is the abelian theory with the moduli space of vacua described by the algebraic integrable system \WitDonagi, an elliptic Calogero-Moser system, which can also be described \gntd,\nekhol\
as a degenerate case of the Hitchin
system \hitchin. The classical elliptic Calogero-Moser system
describes the system of particles
$q_{1}, q_{2}, \ldots , q_{N}$
on  a circle, interacting via a pair-wise potential
$$
U = m^{2} \sum_{i,j=1}^{N} \, {\wp} \left( q_{i} - q_{j} \right)
$$
which is doubly periodic, with the periods $1$ and ${\CT}$, ${\rm Im}{\CT} > 0$, where we use the elliptic modulus defined by the gauge couplings \fdcplng. The classical motion of that system is mapped to the constant velocity motion on the Jacobian variety of the spectral curve,
\eqn\spctrlcrv{{\Det}_{N\times N}\, \left( {\bf \Phi}(z) - {\l} \right) = 0}
where
\eqn\laxopell{{\bf \Phi}_{ij}(z) =
p_{i} {\d}_{ij} + m {{\Theta}_{1} ( z + q_{i} - q_{j} ) {\Theta}_{1}^{\prime}(0) \over
{\Theta}_{1}(z) {\Theta}_{1} (q_{i} - q_{j})} (1 - {\d}_{ij}) }
This family of curves encodes \niksw, \nikokounkov\ the low-energy effective action of the ${\CN}=2^{*}$ theory with the
mass of the hypermultiplet equal to $m$.
The prepotential ${\CF}$ depends on the vacuum expectation values 
$\langle \phi \rangle = {\rm diag} ( a_{1}, \ldots , a_{N} )$
of the scalars in the vector multiplet of the $U(N)$ gauge group, and on $m$ and $\tau$:
\eqn\effprep{{\CF}( a ; m ,{\CT}) =
{\CF}^{\rm pert} ( a ; m, {\CT} ) +
\sum_{k=1}^{\infty} e^{2N k {\pi}i \CT}
{\CF}_{k} (a ; m) \ ,}
where ($a_{ij} = a_{i} - a_{j}$):
\eqn\pertpre{\eqalign{& {\CF}^{\rm pert}( a ; m, {\CT} ) = {{\CT}\over 2} \sum_{i=1}^{N} a_{i}^{2} + 
{3N^{2} m^{2}\over 4} +  {1\over 4} \sum_{i,j=1}^{N}
\left[ a_{ij}^{2}\,
 {\rm log} \left( a_{ij} \right)  -
\left( a_{ij}  +  m\right)^{2}\,
 {\rm log} \left( a_{ij} + m\right) \right] \cr}}
The terms ${\CF}_{k}(a ; m)$
come from the charge $k$ instantons and can be computed for any $k$ using localization techniques \niksw:
\eqn\instcrr{{\CF}_{1}( a; m) = m^{2}
\sum_{i=1}^{N} \prod_{j\neq i} \left( 1 - {m^{2}\over ( a_{ij} )^{2}} \right)\, , \ \rm etc. }{}Now let us apply the same procedure to the ${\CN}=2^{*}$ theory, i.e. let us compactify the theory on a two-sphere with the partial twist. Actually, the theory with adjoint hypermultiplet can be twisted in many ways. Indeed, we have an extra $U(1)$ symmetry under which the complex scalars
$B_{1}, B_{2}$ in the adjoint hypermultiplet have charges $+1, -1$. By embedding $SO(2)_{{\bS}^{2}}$ into this $U(1)$ we shall assign the additional Lorentz  spins to the bosons and fermions in the
hypermultiplet.

The two dimensional twisted superpotential now
contains, in addition to the terms   \wsupmn, the terms coming from the extra twist of the matter fields (we identify $a^{i} = {\s}_{i}$):
\eqn\wsupmnone{{\tilde W}^{\rm eff}({\s}; m,{\CT}) = 2{{\p}{\CF} ( {\s}; m, {\CT} ) \over {\p}m} + \sum_{i=1}^{r} \left( {\bm}_{i} {{\p\CF ( {\s}; m, {\CT} )}\over {\p}{\s}_{i}} + {\bn}_{i} \sigma^{i} \right)}
and vacuum equation is defined with this and \main. We note that the perturbative limit of the   \wsupmnone\ gives the  twisted effective superpotential of the Yang-Mills-Higgs theory of \higgs, \gerasimovshatashvili, \gstwo\ (the example \adjvv\ of the Hitchin theory  above). This is not surprising since in the trivial instanton sector the reduction on ${\bS}^2$ of the four dimensional ${\CN}=2^{*}$ theory gives the two dimensional ${\CN}=2^{*}$ theory. We see here that the four dimensional instation corrections give a modular-covariant deformation of the effective twisted superpotential, and a modular-covariant deformation of the Bethe equations of the non-linear Schr\"odiner system. 

This is a very interesting phenomenon which needs further investigation, see \bass\ for details.

\subsubsec{Higher energies}

Another very exciting direction
of research involves attempting to lift the correspondence between the quantum integrable system and the gauge theory beyond the vacuum sector of the latter. It is conceivable that the
Yangian, quantum affine, or elliptic quantum algebra symmetry of the vacuum sector are the symmetries of the full quantum field theory.
Note that in the two and three dimensional cases
these algebras do not, in general, preserve
the number of colors. We thus see
a novel kind of symmetry of a gauge theory emerging. When the gauge theories are
imbedded in string theory via e.g. a $D$-brane
construction, the change of the rank of the gauge group looks less drastic, as it corresponds
to bringing some branes from infinity or sending them away, see \bass\ for details. 

\footatend\vfill\supereject\immediate\closeout\rfile\writestoppt
\baselineskip=14pt\centerline{{\bf References}}\bigskip{\frenchspacing%
\parindent=20pt\escapechar=` \input refs.tmp\vfill\eject}\nonfrenchspacing
\bye